\documentclass[aps,physrev,twocolumn,groupedaddress]{revtex4-2}

\usepackage{amsmath}
\usepackage{braket}
\usepackage{amsfonts}
\usepackage{amssymb}
\usepackage{tikz}
\usepackage{lipsum}
\usepackage{enumitem}
\usepackage{verbatim}
\usepackage{hyperref}

\DeclareMathOperator{\Tr}{Tr}

\usepackage{placeins}

\usepackage{xcolor}
\definecolor{darkmagenta}{rgb}{0.545, 0.0, 0.545}
\definecolor{violet}{rgb}{0.933, 0.510, 0.933} 
\hypersetup{colorlinks=true, linkcolor=darkmagenta, citecolor=violet}

\begin{document}

\title{The iSWAP gate with polar molecules: Robustness criteria for entangling operations}

\author{Matteo Bergonzoni}

\email{bergonzoni@unistra.fr}
\author{Sven Jandura}

\author{Guido Pupillo}
\email{pupillo@unistra.fr}
\affiliation{University of Strasbourg and CNRS, CESQ and ISIS (UMR 7006), 67000 Strasbourg, France}

\date{\today}

\begin{abstract}

Ultracold polar molecules trapped in optical lattices or tweezer arrays are an emerging platform for quantum information processing and quantum simulation, thanks to their rich internal structure and long-range dipolar interactions. Recent experimental breakthroughs have enabled precise control over individual molecules, paving the way for implementing two-qubit quantum gates, based on the iSWAP gate. A key challenge is however the sensitivity to variations of the dipole-dipole interaction strength -- stemming from motion of the molecules  and uncertainty on the precise positioning of external confining potentials --  that limits current gate fidelities. To address this, we develop a quantum optimal control framework, based on a perturbative approach, to design gates that are robust with respect to quasi-static deviations of either system Hamiltonian parameters, external control parameters, or both, and provide criteria to evaluate a priori whether a gate can be made robust for a given control Hamiltonian. By applying these criteria to exchange-coupled qubits, as polar molecules, we demonstrate that robustness cannot be achieved with global controls only, but can be attained by breaking the exchange symmetry through local controls, such as a local detuning. We determine the robust time-optimal solution for realizing an iSWAP gate, show that the control pulses can be designed to be smooth functions, and achieve theoretical gate fidelities compatible with error correction using reasonable experimental parameters. Additionally, we show that certain entangled state preparations, such as Bell states, can be made robust even with global controls only. We demonstrate that, under the adiabatic approximation -- i.e. where molecular motion occurs on timescales comparable to, or faster, than that of the exchange interaction -- the noise arising from thermal motion can be effectively treated as quasi-static variations of Hamiltonian parameters. This allows us to extend our treatment to the concrete experimental case of polar molecules.

\end{abstract}

\maketitle

\section{Introduction}

Ultracold molecules trapped in optical lattices or tweezer arrays are a leading platform for the simulation of many-body quantum systems \cite{Ni_2008, Baranov_2012, Quemener_2012, Takekoshi_2014, Molony_2014, Barry_2014, Park_2015, Molony_2016, Guo_2016, Rvachov_2017, Truppe_2017, Anderegg_2017, Seesselberg_2018, Collopy_2018, Yang_2019b, Hu_2019, Voges_2020, Vilas_2020, Kaufman_2021, Cairncross_2021, Stevenson_2023, Ruttley_2024, Picard_2024, Cornish_2024, Holland_2024}. In recent years, significant advances have been made in the preparation of individual polar molecules in optical traps \cite{Ni_2008, Danzl_2008, Lang_2008, Anderegg_2019, Voges_2020, Zhang_2022, Guttridge_2023}, controlling and reading out single quantum states \cite{Anderegg_2019, Cairncross_2021, Rosenberg_2022, Ruttley_2023}, and extending hyperfine and rotational coherence \cite{Park_2017, Gregory_2021, Burchesky_2021, Lin_2022, Park_2023}. These advances now unlock new opportunities to exploit the unique properties of molecules for quantum information applications. The rich internal structure of molecules provides many stable states with long coherence times. Moreover, the long-range, tunable dipolar interaction implements an exchange Hamiltonian, and thus it naturally enables the entanglement of molecular pairs and the implementation of entangling gates, e.g. the iSWAP gate. There have already been several proposals on how to realize high-fidelity two-qubit entangling gates with molecules \cite{DeMille_2002, Yelin_2006, Jing_2013, Park_2017, Ni_2018, Hughes_2020, Gregory_2021,  Holland_2023, Bao_2023, Picard_2024_arXiv, Ruttley_2025} and the state-of-the-art fidelities for the preparation of individual Bell states and of the iSWAP gate have reached, after post-selection to remove SPAM errors,  $F=0.976$ \cite{Ruttley_2025} and $F=0.92$ \cite{Picard_2024_arXiv}, respectively. While promising, these results are  still more than an order of magnitude below what is possible today with leading quantum computing platforms such as neutral atoms, ions or superconducting qubits \cite{Ballance_2016, Ding_2023, Peper_2025}. One key fundamental challenge that these protocols have to face is sensitivity to variations of the dipole-dipole interaction strength at the core of the iSWAP gate, induced by, e.g., the relative motion of the molecules \cite{Holland_2023, Bao_2023, Picard_2024_arXiv, Cornish_2024}, or spin-motion coupling \cite{Picard_2024_arXiv}.\\

Fidelities of two- and multi-qubit gates are often limited by uncertainties in the system Hamiltonian parameters of a given quantum computing platform, noise in the experimental control fields, and coupling to the environment. 
In order to increase gate fidelities, a variety of quantum optimal control methods -- corresponding to a set of techniques designed to perform desired operations on quantum systems using time-varying control fields \cite{Glaser_2015, Werschnik_2007, Wilhelm_2020, Koch_2022, Ansel_2024} -- have
been successfully developed and utilized in recent years, including with superconducting qubits \cite{Egger_2013, Kelly_2014, Huang_2014, Theis_2018, Werninghaus_2021}, trapped ions \cite{Nebendahl_2009, Choi_2014, Bentley_2020}, neutral atoms \cite{Goerz_2011, Muller_2011, Goerz_2014, Anderson_2015,  Larrouy_2020, Jandura_2022, Ma_2023, Evered_2023}, color centers in diamond \cite{Konzelmann_2018, Dong_2021, Chakraborty_2022}, and spin qubits \cite{Yang_2019, Cerfontaine_2020, Cerfontain_2020b}.
Several strategies have been proposed to realize robust control protocols, including: the development of  different-phase contiguous pulse sequences
\cite{Tycko_1983, Levitt_1986, Wimperis_1994, Wang_2012, Owrutsky_2012, Daems_2013, Ichikawa_2013, Bando_2013}, controller synthesis problem for linear stochastic systems \cite{James_2008, dekeijzer_2024}, 
extension of the Pontryagin maximum principle to control optimization in the presence of noise  \cite{Boscain_2021, Damme_2017, Dridi_2020, Koswara_2021, Ansel_2024}, adiabatic rapid passage \cite{Mitra_2020, Li_2023, Chanda_2023}, supervised learning models \cite{Wu_2019, Shi_2024}, so-called sampling or ensemble method 
\cite{Khaneja_2005, Chen_2014, Goerz_2014, Hughes_2020, Rembold_2020, Giudici_2024}, 
universal robust quantum control \cite{Poggi_2024}, and the derivative or perturbative method \cite{Propson_2022, Jandura_2023, Madhav_2023}. Despite these numerous advances, general criteria for determining a priori whether a gate can be made robust against a specific noise model, given the available control fields, are largely missing.\\

In this work, we propose a general theoretical framework for designing {\it robust} quantum gates using a quantum optimal control method based on the derivative -- or perturbative -- method furnishing also criteria to determine \textit{a priori} if a gate can be made robust or not for a given control Hamiltonian. Our method is based on the assumption that the noise is a quasi-static fluctuation of a Hamiltonian parameter. Surprisingly, we demonstrate both theoretically and through numerical simulations that the adopted robustness requirement remains effective even under weak coupling to a noisy environment -- such as a thermal bath -- within the adiabatic approximation: when the time scale over which the environment evolves is small compered to the time scale over which the state of the system varies appreciably, it is possible to trace out the environment and effectively model its influence as a quasi-static noise parameter in the system Hamiltonian. We apply and test these criteria in the specific case of a quantum gate for two-qubits  that are coupled via an exchange Hamiltonian, aiming to reduce sensitivity to variations in the interaction strength. This is the situation of two trapped polar molecules interacting via dipole-dipole interaction and implementing an iSWAP gate, where static deviations of the trapping potentials or the motion of the molecules are a major source of error. We demonstrate that  robustness of the iSWAP gate cannot be achieved only using a global control that drives symmetrically the qubits. However, robustness can be achieved by breaking the exchange symmetry by adding local controls, e.g. detunings in local control  or confining fields. 
We provide a pulse-level protocol that realizes fast iSWAP gates that are robust to quasi-static variations of the strength of dipole-dipole interactions, thus also to molecular motion. We focus on time optimal and robust pulses that are smooth functions of control parameters -- and thus probably easier to implement in experiments -- achieving a fidelity $F>0.9996$ even in the presence of variations of 10\% in the interaction strength. Interestingly, we also demonstrate that certain entangled state preparations, such as Bell states, can still be achieved robustly with global controls. We provide realistic scenarios for the realization of the protocols for fast and robust quantum gates in current and near-future experiments with cold polar molecules.\\

There are two main implications of this work: First, we demonstrate that high-fidelity iSWAP gates should be achievable with current or near-future technology with polar molecules. For the considered noise sources, we achieve theoretical gate fidelities compatible with error correction using reasonable experimental parameters for the iSWAP gate, for the first time. It should be emphasized that the considered noise sources -- such as axial motion in the traps -- are precisely those identified by experiments as the dominant source of errors. Second, many of the considerations of general nature on the controllability of qubits and the realization of robust gates can be directly translated to any other platforms. In particular, the robust iSWAP gate presented here for polar molecules could be implemented in neutral atom quantum processors, where the dipole-dipole interactions couples different excited Rydberg states. \\

The paper is structured as follows. In Sec.~\ref{section_1} we introduce the general framework for a quantum optimal control problem involving robustness to variations in some physical parameter. We also present a set of sufficient conditions to assess \textit{a priori} whether it is possible to achieve a robust gate given a specific control Hamiltonian. We demonstrate that the adopted robustness definition accounts not only for parametric fluctuations, but also for noise arising from coupling to a environment, e.g. molecular motion. In Sec.~\ref{section_2}, we apply these general tools to the specific case of an exchange Hamiltonian implementing an iSWAP gate. We find out that the previously established criteria rule out the possibility of a global control Hamiltonian making the gate robust. However, we propose a protocol to work around this issue by breaking the exchange symmetry. In Sec.~\ref{sect:physical_realizations}, we present two possible physical realizations of the iSWAP gate, such as polar molecules and Rydberg atoms interacting via dipole-dipole interaction. We also analyze how molecular motion and motion-rotation coupling affect the interaction strength. In Sec.~\ref{section_3}, we address the topic of numerical methods, such as GRAPE and Chebyshev decomposition, to construct robust pulses that implement an iSWAP gate. In Sec.~\ref{sect:results}, we show the results of the numerical optimization. In particular, we obtain a smooth pulse that has a fidelity $F>0.9996$ even in the presence of variations of 10\% in the interaction strength. We also propose a protocol for the robust preparation of a Bell state, relying solely on global controls.\\
All pulses found in this work are available at Ref. \cite{data_2025}.

\begin{figure*}
    \centering
    \includegraphics[width=0.98\linewidth]{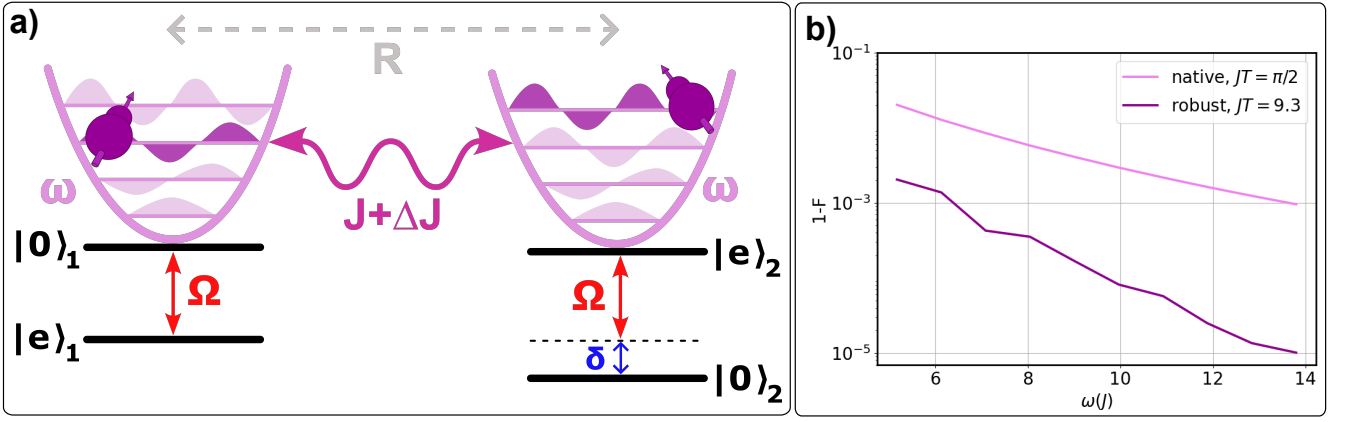}
    \caption{\textbf{a)} Schematic representation of trapped polar molecules interacting via dipole-dipole interaction and implementing an iSWAP gate. The states $\ket{0}_j$ and $\ket{e}_j$ are rotational states of the molecule $j$, coupled by a spin-exchange Hamiltonian as in Eq.~\eqref{H_0}, with coupling strength $J \propto R^{-3}$, where $R$ is the intermolecular distance. The trapping potentials can be modeled as harmonic oscillators of frequency $\omega$, and the finite width of the motional wavefunction within each tweezer generates an uncertainty $\Delta R$ in the distance $R$, leading to an associated uncertainty $\Delta J$ in the interaction strength $J$. To reduce the sensitivity to variations $\Delta J$ in the interaction strength, an external global microwave drive $\Omega(t)$ and a local detuning $\delta$ are introduced, as in Eq.~\eqref{H_c2}. \textbf{b)} Infidelity $1 - F$ of the non-robust native gate (light line) and the Chebyshev robust pulse shown in Fig.~\hyperref[fig:big_pic]{\ref{fig:big_pic}(f)} (dark line), plotted as a function of the trapping frequency $\omega$, expressed in units of the coupling strength $J$. The simulation is performed by considering quantum motion in the axial trapping harmonic oscillator -- the dominant source of decoherence [see Sec.~\ref{sec:mot-rot_coupling}] -- as described in Eq.~\eqref{H_0_delta_R}. The robust pulse was optimized under the assumption of quasi-static fluctuations in the coupling $\Delta J$, but it retains its robustness even in the presence of motion [see Sec.~\ref{sec:noise-coupling_env}], yielding an improvement in infidelity of one to two orders of magnitude. The improvement is greater for larger trapping frequencies $\omega$, as the adiabatic approximation regime -- i.e., $J \gg \omega$ -- is more accurately satisfied in this case. The simulation is performed with 7 motional states.}
    \label{fig:motion_omega}
\end{figure*}

\section{\label{section_1}Theoretical framework}

In this section we introduce the theoretical tools used in this work. In Sec.~\ref{sect:qoc_problem}, we introduce the general framework for a quantum optimal control problem involving robustness to quasi-static variations in the Hamiltonian or control parameters. In Sec.~\ref{sect:robustness}, using a derivative-based method, we define a measure to quantify the robustness of a gate. In Sec.~\ref{sect:criteria}, we present a set of sufficient conditions to evaluate \textit{a priori} whether it is possible to make a gate robust for a given system and control Hamiltonian. In Sec.~\ref{sec:noise-coupling_env}, we demonstrate that, under adiabatic approximation, the noise arising from a coupling to the environment, e.g. thermal motion, can be removed with the same method of quasi-static variations of Hamiltonian parameters.

\subsection{\label{sect:qoc_problem}The quantum optimal control problem}

Consider a generic quantum system whose internal dynamics is governed by the Hamiltonian $H_0(\vec{J})$ that depends on a set of physical parameters $\vec{J}$. This Hamiltonian can be considered time-independent, up to a frame transformation. Due to experimental imperfections, the parameters $\vec{J}$ may be subject to quasi-static noise $\Delta \vec{J}$. Quasi-static noise refers to unknown and random variations in the values of $\vec{J}$, that can vary from one experiment to the next, but they remain constant during a single experiment. A more detailed description of the noise model for molecules is presented in Sec.~\ref{sec:mot-rot_coupling}.\\

The quantum optimal control problem under study concerns the implementation of a target unitary operator $U_T$, i.e. a quantum gate, by acting on the system with a control Hamiltonian $H_c(\vec{u}(t))$. The latter depends on a set of control fields $\vec{u}(t)$, which can be time-dependent. Control fields may be affected by experimental imperfections described by the uncertainty $\Delta \vec{u}$ in the control parameters, which is assumed to be quasi-static. The total Hamiltonian $H$, including both the system Hamiltonian $H_0$ and the control Hamiltonian $H_c$, reads 
\begin{equation}
    \label{H_total}
    H(\vec{J},\vec{u}(t)) = H_0(\vec{J}) + H_c(\vec{u}(t)).
\end{equation}
In the following, ``robustness" with respect to noise on a given system or control parameter refers to the realization of a {\it final quantum state} that is insensitive to first-order variations of that parameter, in a perturbative sense. The quantum optimal control problem can then be rephrased as follows: Can a set of optimal parameters $\vec{u}(t)$ be found, such that the evolution operator of the total Hamiltonian $H$, i.e. Eq.~\eqref{H_total}, reproduces the desired target gate $U_T$ up to  order $o(|\Delta\vec{J}|^2)$ and $o(|\Delta\vec{u}|^2)$ in perturbation theory? In mathematical form this question reads
\begin{align}
    \nonumber
    \mathcal{T} & \exp\Bigg\{-i \int_0^{T} dt H(\vec{J} + \Delta\vec{J}, \vec{u}(t) + \Delta\vec{u})\Bigg\} =\\
    \label{problem0}
    & = U_T + o(|\Delta \vec{J}|^2) + o(|\Delta \vec{u}|^2),
\end{align}
where $\mathcal{T}$ denotes the usual time ordering and $T$ is the total time duration of the gate. In all this work we set $\hbar=1$.\\

We note that the only difference between physical parameters $\vec{J}$ and control parameters $\vec{u}(t)$ is that the former describe the internal evolution of the system, and thus usually cannot be controlled, whereas the latter can be adjusted and modified over time to achieve the desired gate. For simplicity, the following discussion focuses on noise on the  physical parameters $\vec{J}$ only, though it can be straightforwardly extended to noise in the controls $\vec{u}(t)$ \cite{Hughes_2020, Jandura_2023}. We point out that the following discussion on robustness also applies to the case of a system coupled to a noisy environment, as discussed in detail in Sec.~\ref{sec:noise-coupling_env}.

\subsection{\label{sect:robustness}Robustness of a quantum gate}

This section introduces a precise definition and measure of robustness, based on a perturbative approach. This approach is useful when the Hamiltonian of the system under study is known, as is the case for platforms such as polar molecules and neutral atoms. For simplicity, we assume that noise affects only one of the system parameters, say $J$ -- it is straightforward to generalize the discussion to the case where multiple parameters are affected by noise.\\ 

Our goal is to eliminate the first-order effects of noise $\Delta J$ on $J$, using standard perturbation theory. The total Hamiltonian $H$ of Eq.~\eqref{H_total} can then be expanded in powers of $\Delta J$ in a perturbative approach as
\begin{equation}
    \label{H_decomposition}
    H(t) = H^{(0)}(t) + \Delta J H^{(1)} + o(\Delta J^2),
\end{equation}
where $H^{(0)}$ and $H^{(1)}$ are the zeroth- and first-order Hamiltonians in the perturbation $\Delta J$, respectively. The perturbative approach applied to a generic state $\ket{\psi(t)}$ describing the system at time $t$ reads
\begin{equation}
    \label{psi_decomposition}
    \ket{\psi(t)} = \ket{\psi^{(0)}(t)} + \Delta J \ket{\psi^{(1)}(t)} + o(\Delta J^2).
\end{equation}
The time-evolution of the state is governed by the Schr\"odinger equation
\begin{equation}
    \label{sch_eq}
    \frac{d\ket{\psi(t)}}{dt} = -i H(t) \ket{\psi(t)}.
\end{equation}
Taking into account the perturbative expansion of Hamiltonian $H$ of Eq.~\eqref{H_decomposition} and of the state $\ket{\psi}$ of Eq.~\eqref{psi_decomposition} in terms of  $\Delta J$, the Schr\"odinger equation \eqref{sch_eq} can be written in matrix form as
\begin{equation} 
    \label{sch_eq2}
        i\frac{d}{dt}
        \begin{pmatrix}
            \ket{\psi^{(0)}(t)}\\
            \ket{\psi^{(1)}(t)}
        \end{pmatrix}
         =
         \begin{pmatrix}
             H^{(0)}(t) & 0 \\
             H^{(1)} & H^{(0)}(t)
         \end{pmatrix}
         \begin{pmatrix}
            \ket{\psi^{(0)}(t)}\\
            \ket{\psi^{(1)}(t)}
        \end{pmatrix}.
\end{equation}
We assume that the system is initially prepared in a state $\ket{q}\in Q$, where $Q$ is a basis set of the Hilbert space of the system, e.g. the computational two-qubit basis $\{\ket{00}, \ket{01}, \ket{10}, \ket{11}\}$. One can further assume that the first-order state $\ket{\psi^{(1)}(0)}$ at $t=0$ is zero. In this case the initial conditions read
\begin{equation}
\label{initial_conditions}
    \begin{pmatrix}
        \ket{\psi^{(0)}(0)}\\
        \ket{\psi^{(1)}(0)}
    \end{pmatrix}
    =
    \begin{pmatrix}
        \ket{q}\\
        0
    \end{pmatrix}.
\end{equation}
Given these initial conditions one can write a formal solution to the Schr\"odinger equation \eqref{sch_eq2}. The zeroth-order state at the final time $T$ reads
\begin{equation}
    \label{psi_0}
    \ket{\psi_q^{(0)}(T)} = U^{(0)}(0,T) \ket{q},
\end{equation}
where $\ket{\psi_q^{(0)}(t)}$ is obtained at time $t$ by evolving $\ket{q}$, with $U^{(0)}(0,t)$ the evolution operator of the zeroth-order Hamiltonian $H^{(0)}$
\begin{equation}
    U^{(0)}(0,t) = \mathcal{T} \exp\left\{-i\int_0^t dt' H^{(0)}(t')\right\}.
\end{equation}
The corresponding first-order state reads instead
\begin{equation}
    \label{psi_1}
    \ket{\psi_q^{(1)}(T)} = -i \int_0^T dt U^{(0)}(t,T) H^{(1)} U^{(0)}(0,t) \ket{q}.
\end{equation}
From the definition of a robust gate in Eq.~\eqref{problem0}, the robustness $\mathcal{R}$ with respect to the parameter $J$ can be defined as the sum of norms of first-order states $\ket{\psi_q^{(1)}(T)}$ from Eq.~\eqref{psi_1} as
\begin{equation}
    \label{R}
    \mathcal{R} = \sum_{q\in Q} \braket{\psi_q^{(1)}(T)|\psi_q^{(1)}(T)}.
\end{equation}
A robust gate will correspond to a gate for which $\mathcal{R}=0$, i.e. the norm of the first-order state $\ket{\psi_q^{(1)}(T)}$ at the end of the pulse ($t=T$) vanishes for any possible initial basis state $\ket{q}$ \cite{Jandura_2023}.\\

In principle, this discussion can be extended to include the norm of higher-order states in the definition of robustness in Eq.~\eqref{R}, e.g. for the second-order term $\sum_{q} \braket{\psi_q^{(2)}(T)|\psi_q^{(2)}(T)}$. By doing so, one introduces more stringent constraints that the controls must satisfy. As a result, when transitioning to numerical optimization, one typically obtains pulses that, while more robust to noise, tend to be longer and more complex (e.g. not smooth functions of time), thus probably more difficult to implement in experiments than those obtained from Eq.~\eqref{R}. In the following, we thus focus on the latter only [see an example in Sec.~\ref{sect:grape}]. \\

In App. \ref{sect:ext_robustness}, we discuss a simple generalization of the robustness criterion $\mathcal{R} = 0$, allowing the first-order states $\ket{\psi_q^{(1)}(T)}$ to have a nonzero norm while constraining them to be parallel to the desired states $\ket{\psi_q^{(0)}(T)}$, thereby defining an extended robustness criterion $\mathcal{R}_e=0$ in Eq.~\eqref{R_e}. This generalization is not essential for the remainder of this work, thus we present it in App. \ref{sect:ext_robustness}.

\subsection{\label{sect:criteria}Robustness criteria 
for quantum gates}

In many situations, it may be advantageous to know \textit{a priori} whether it is in principle possible (or not) to construct a robust quantum gate with respect to given uncertainties in system and control parameters. 
In this section, we provide general criteria to reply to this question, for the case when both the system and control Hamiltonians are known. For this, we insert Eq.~\eqref{psi_1} for $\ket{\psi_q^{(1)}(T)}$ into the expression for the robustness $\mathcal{R}$ [Eq.~\eqref{R}] obtaining 
\begin{align}
    \mathcal{R} = & \sum_{q\in Q} \braket{\psi_q^{(1)}(T)|\psi_q^{(1)}(T)}\\
    \nonumber
    = & \int_0^T dt \int_0^T dt' \sum_{q\in Q} \bra{q} U^{(0)}(0,t)^\dag H^{(1)} U^{(0)}(t,T)^\dag\cdot\\
    & \cdot U^{(0)}(t',T) H^{(1)} U^{(0)}(0,t')\ket{q}\\
    \label{comp_2}
    = & \int_0^T dt \int_0^T dt' \Tr\Big\{ H^{(1)} U^{(0)}(t',t) H^{(1)} U^{(0)}(t,t')\Big\},
\end{align}
where the fact that $Q$ is a basis is used to express the sum as a trace, and the cyclic property of the trace is applied to permute the evolution operators.
From Eq.~\eqref{comp_2}, we identify three criteria for which the quantum gate cannot be made robust to noise in $J$ using external control parameters $\vec{u}(t)$. These are
\begin{enumerate}[label=(\roman*)]
    \item \label{criterion_i} \textbf{Commuting Hamiltonians} $\mathbf{[H^{(1)}, H^{(0)}(t)] = 0}$. If the zeroth and the first-order Hamiltonian commute at any time, then also the zeroth-order evolution operator $U^{(0)}$ commutes with $H^{(1)}$. This implies that $\mathcal{R} = T^2\Tr\{{H^{(1)}}^2\}>0$ because the squared of a matrix is always positive semi-definite and its trace is non-negative, zero only in the trivial case $H^{(1)}=0$.
    \item \label{criterion_ii} \textbf{Positive semi-definite Hamiltonian }$\mathbf{H^{(1)}\succeq0}$. Consider the following two algebraic properties of positive semi-definite matrices \cite{Horn_Johnson_2012}
    \begin{itemize}
        \item If $A$ is a positive semi-definite matrix then for every matrix $B$ it holds that $B^\dag A B$ is a positive semi-definite matrix as well.
        \item If $A$ and $B$ are two positive semi-definite matrices, then $\Tr\{AB\}\geq0$.
    \end{itemize}
    From the first property, it follows that the expression $U^{(0)}(t',t) H^{(1)} U^{(0)}(t',t)^\dag$ is positive semi-definite. From the second property, it follows that the trace of $H^{(1)}U^{(0)}(t',t) H^{(1)} U^{(0)}(t',t)^\dag$ is non-negative $\forall t,t'\in[0,T]$, thus the integral in Eq.~\eqref{comp_2} is positive and $\mathcal{R}>0$. In principle $\mathcal{R}$ could be zero, but this would require the operator inside the trace be equal to zero $\forall \enspace t,t'\in[0,T]$, which only occurs in the trivial situation $H^{(1)}=0$.
    \item \label{criterion_iii} \textbf{Combination of commuting and positive semi-definite cases} $\mathbf{H^{(1)}=PC}$ \textbf{s.t.} $\mathbf{P\succeq0,}$ $\mathbf{[C,H^{(0)}(t)]=0,}$ $\mathbf{ C=C^\dag}$. The first-order Hamiltonian can be decomposed in the product of two matrices, $P$ and $C$, where $P$ is positive semi-definite, and $C$ commutes with $H^{(0)}$ and is Hermitian. Given these assumptions, Eq.~\eqref{comp_2} for the robustness can be rewritten as 
    \begin{align}
        \nonumber
        \mathcal{R} = & \int_0^T dt \int_0^T dt' \Tr\Big\{ P \left(U^{(0)}(t',t) C\right) \cdot\\
        & \cdot P \left(U^{(0)}(t',t)C\right)^\dag\Big\}.
    \end{align}
    Applying the two algebraic properties given above in criterion \ref{criterion_ii}, it is straightforward to verify that the robustness has to be greater than zero, $\mathcal{R}>0$. Again, it is zero only in the trivial case in which $H^{(1)}=0$.
\end{enumerate}
These three criteria enable one to determine \textit{a priori} whether a quantum gate can be made robust against certain noisy parameters, for a given set of control fields. Criterion \ref{criterion_i} formalizes the well-known intuitive idea that a control Hamiltonian that commutes with the system Hamiltonian cannot reduce sensitivity to noise, as is the case with, e.g., dynamical detuning \cite{Viola_1998, Ng_2011}. Concerning criterion \ref{criterion_ii}, we note that Ref. \cite{Jandura_2023} demonstrated the impossibility of realizing a CZ gate that is robust to detuning variations -- in the context of quantum gates for Rydberg atoms -- by using  arguments that can be recast into our criterion \ref{criterion_ii} given above. To our knowledge, the criterion \ref{criterion_iii} has never been explicitly mentioned in the literature. Its relevance is demonstrated in a concrete example in Sec.~\ref{sec:robust?_no_sorry} by the application of criterion \ref{criterion_iii} to the case of a symmetric exchange interaction between polar molecules, where {\it global} control fields are demonstrated to be insufficient to realize a iSWAP gate that is robust with respect to fluctuations of the relative position, thus the interaction strength.\\

We note that it is, in principle, possible to realize a robust quantum gate even if the criteria \ref{criterion_i} and \ref{criterion_iii} are satisfied for a given set of controls $H_c$, provided it is possible to modify the controls -- e.g. in experiments. Essentially, this involves altering the zeroth-order Hamiltonian $H^{(0)}$ and its commutation relations with the first-order Hamiltonian $H^{(1)}$. In Sec.~\ref{sect:solution} we provide a concrete example for the case of the iSWAP gate with polar molecules, which can be achieved using {\it local} control fields, as opposed to global ones.\\

Interestingly, we note that the criteria \ref{criterion_i} - \ref{criterion_iii} discussed above apply only to the realization of robust quantum gates $U_T$. However, if one is instead interested in robustly preparing a {\it given state} only, such as a maximally entangled Bell state, starting from a given initial condition (i.e. without concern for the robustness of the evolution of all other states in the Hilbert space), the situation might be different: We discuss in Sec.~\ref{sect:state_preparation} that a Bell-state can in fact be realized robustly with polar molecules using global fields only, even if the criteria \ref{criterion_i} - \ref{criterion_iii} are satisfied.

\subsection{\label{sec:noise-coupling_env}Noise from the interaction with the environment}

The results of the sections above hold in the presence of a small unknown quasi-static fluctuation $\Delta J$ of a physical parameter $J$ with first-order noise Hamiltonian $H^{(1)}$. The generalization of robustness to an open quantum system $S$ coupled to an environment $E$ is not straightforward. However, in this section it is shown that making a quantum operation robust to coupling with a noisy environment is equivalent to making the same quantum operation robust to quasi-static parameter uncertainty, choosing an appropriate noise Hamiltonian $H^{(1)}$, under the adiabatic assumption. This allows us to extend our discussion of Sec.~\ref{sect:criteria} to the case of noise arising from the coupling to an environment, e.g. thermal bath.

\subsubsection{Parameter uncertainty}

Consider an isolated quantum system $S$ described by the Hilbert space $\mathcal{H}_S$ and with a noisy parameter $J$. The dynamics of the system is governed by the Hamiltonian $H$ in Eq.~\eqref{H_decomposition}. It has already been discussed how the state $\ket{\psi(t)}$ in Eq.~\eqref{psi_decomposition} can be expanded perturbatively in powers of the small quasi-static noise $\Delta J$. The evolution of the state is governed by Schr\"odinger equation in Eq.~\eqref{sch_eq2}, with the initial conditions in Eq.~\eqref{initial_conditions}. The solution of Schr\"odinger equation for the zeroth-order state $\ket{\psi^{(0)}_q(T)}$ at the final time $T$ is in Eq.~\eqref{psi_0}, while the final first-order state $\ket{\psi^{(1)}_q(T)}$ is in  Eq.~\eqref{psi_1}. In the following discussion it will be useful to work with the density matrix formalism, thus the perturbative expansion is introduced also for the density matrix $\rho_q(t)=\ket{\psi_q(t)}\bra{\psi_q(t)}$ as
\begin{equation}
    \label{rho_decomposition}
    \rho_q(t) = \rho^{(0)}_q(t)+\Delta J\rho^{(1)}_q(t)+\Delta J^2\rho^{(2)}_q(t),
\end{equation}
where the second-order term in $\Delta J$ is kept to account for for all first-order terms in the state $\ket{\psi^{(1)}(t)}$, such that 
\begin{align}
        \label{rho_1}
        \rho^{(0)}_q(t)  &=\ket{\psi^{(0)}(t)}\bra{\psi^{(0)}(t)},\\
        \label{rho_2}
        \rho^{(1)}_q(t) &=\ket{\psi^{(0)}(t)}\bra{\psi^{(1)}(t)} + \text{h.c.},\\
        \label{rho_3}
        \rho^{(2)}_q(t) &=\ket{\psi^{(1)}(t)}\bra{\psi^{(1)}(t)}.
\end{align}
Using the Eqs.~\eqref{rho_1}-\eqref{rho_3} for the different perturbative orders of the density matrix $\rho_q(t)$, and the Eqs.~\eqref{psi_0} and \eqref{psi_1} for the different perturbative orders of the state $\ket{\psi_q(t)}$, it can be shown that
\begin{align}
        \label{rho_1_bis}
        \rho^{(0)}_q(T)  = & U^{(0)}(0,T)\ket{q}\bra{q}U^{(0)}(T,0),\\
        \nonumber
        \rho^{(1)}_q(T) = & -i \int_0^Tdt \Big[ U^{(0)}(0,T)\ket{q}\bra{q}\cdot\\
        \label{rho_2_bis}
        & \cdot U^{(0)}(t,0)H^{(1)}U^{(0)}(T,t)+ \text{h.c.} \Big],\\
        \nonumber
        \rho^{(2)}_q(T) = & -\int_0^Tdt \int_0^Tds U^{(0)}(t,T) H^{(1)} U^{(0)}(0,t)\cdot\\
        \label{rho_3_bis}
        &\cdot\ket{q}\bra{q} U^{(0)}(s,0)H^{(1)}U^{(0)}(T,s).
\end{align}
The definition of robustness $\mathcal{R}=0$ in Eq.~\eqref{R} ensures that the first-order state $\ket{\psi_q^{(1)}(t)}$ has zero norm for all basis state $\ket{q}$. Thus, from Eqs.~\eqref{rho_1}-\eqref{rho_3} it is easy to see that the condition $\mathcal{R}=0$ ensures also that $\rho_q^{(1)}(T)=\rho_q^{(2)}(T)=0$ in Eqs.~\eqref{rho_2_bis} and \eqref{rho_3_bis}.\\
The above equations hold for an isolated system. However, we find that, under the adiabatic assumption, if the system $S$ is coupled to a noisy environment $E$, the condition $\mathcal{R} = 0$ remains sufficient to ensure the robustness of the quantum operation. This is discussed in details in the following section.

\subsubsection{\label{subsec:coupling_noise}Coupling to environment}
\label{sect:coupling_environment}

Consider an open quantum system $S$ coupled to a noisy environment $E$, e.g. a thermal bath, such that the full space is described by the Hilbert space $\mathcal{H}_S\otimes\mathcal{H}_E$. In a  perturbed approach the Hamiltonian of the full space is
\begin{equation}
    \label{H_environment}
    H=H_S^{(0)}\otimes\mathbb{I}_E + \mathbb{I}_S \otimes H_E^{(0)} + \epsilon H^{(1)}_S\otimes H^{(1)}_E,
\end{equation}
where $H_S^{(0)}$ and $H_E^{(0)}$ are the Hamiltonians describing the free evolution of the system $S$ and the environment $E$, respectively, while $H^{(1)}_S\otimes H^{(1)}_E$ is the noisy-interaction Hamiltonian between system and environment, with  a small coupling $\epsilon$. Thus, the full zeroth-order Hamiltonian is $H^{(0)}=H_S\otimes\mathbb{I}_E + \mathbb{I}_S \otimes H_E$ and the full first-order Hamiltonian is $H^{(1)}=H^{(1)}_S\otimes H^{(1)}_E$. The density matrix of the full space $\rho(t)$ can be expanded perturbatively in powers of $\epsilon$ as in Eq.~\eqref{rho_decomposition}. The initial conditions for the full density matrix $\rho(t)$ are
\begin{align}
        \label{rho0(0)}
        \rho_q^{(0)}(0) & = \sum_p b_p\ket{q,p}\bra{q,p},\\
        \rho^{(1)}_q(0) & = 0,\\
        \rho^{(2)}_q(0) & = 0,
\end{align}
with $\ket{q,p}=\ket{q}_S\otimes\ket{p}_E$, where $\{\ket{q}_S\}$ and $\{\ket{p}_E\}$ are a set of basis states for the system $\mathcal{H}_S$ and the environment $\mathcal{H}_E$, respectively. In particular we can assume $\{\ket{p}_E\}$ to be the eigenstates of the zeroth-order Hamiltonian $H_E^{(0)}$ and with corresponding eigenvalues $\{E_p\}$. In the following, $\ket{q}$ always refers to a system $S$ state and $\ket{p}$ always refers to an environment $E$ state. For the initial zeroth-order density matrix $\rho_q^{(0)}(0)$ in Eq.~\eqref{rho0(0)} a thermal state at temperature $T_{\text{temp}}$ is assumed, with $b_p$ the probability distribution of the corresponding $\ket{p}$ states, e.g. Boltzmann factors $b_p\propto \exp(-E_p /k_BT_{\text{temp}})$.\\
Solving Schr\"odinger equation for the states $\ket{\psi_{q,p}(t)}$, it is possible to derive expressions for the full density matrix $\rho(t)$ completely analogous to the ones in Eqs.~\eqref{rho_1_bis}-\eqref{rho_3_bis}, with the only difference that now the operators are acting on the full Hilbert space $\mathcal{H}_S\otimes\mathcal{H}_E$ and not only on that one of the system $S$.\\

We now consider the final first-order density matrix $\rho_q^{(1)}(T)$, as in Eq.~\eqref{rho_2_bis}, and trace out the environment degrees of freedom to obtain the evolution of the system only
\begin{align}
     \rho^{(1)}_{q,S}(T)=&  \text{Tr}_E\left\{ \rho^{(1)}_{q}(T) \right\}\\
    \nonumber
    =& -i \sum_{p,p'} b_p \int_0^T dt  \Bigg[ \bra{p'} U^{(0)}(0,T)\ket{q,p}\bra{q,p}\cdot\\
    &\cdot U^{(0)}(t,0)H^{(1)}U^{(0)}(T,t)\ket{p'}+ \text{h.c.} \Bigg]\\
    \nonumber
    = &  -i \sum_{p} b_p \int_0^T dt \Bigg[ U^{(0)}_S(0,T)\ket{q}\bra{q} U^{(0)}_S(t,0) \cdot \\
    & \cdot H^{(1)}_SU^{(0)}_S(T,t) \cdot \bra{p} H^{(1)}_E\ket{p}+ \text{h.c.} \Bigg]\\
    \nonumber
    = & -i  \int_0^T dt \Bigg[U^{(0)}_S(0,T)\ket{q}\bra{q} U^{(0)}_S(t,0)H^{(1)}_S \cdot \\
    \label{expr_rho1}
    &\cdot U^{(0)}_S(T,t) + \text{h.c.}\Bigg]\left( \sum_{p} b_p\bra{p} H^{(1)}_E\ket{p}\right)
\end{align}
where the facts that $U_E^{(0)}(s,t)\ket{p} = \exp[-iE_p(t-s)]\ket{p}$, and $\braket{p|p'}=\delta_{p,p'}$ are used, since $\ket{p}$ are eigenstates of $H_E^{(0)}$.\\
Comparing the expression for the first-order reduced density matrix $\rho^{(1)}_{q,S}$ in Eq.~\eqref{expr_rho1} with the one in Eq.~\eqref{rho_2_bis} for the first-order density matrix of an isolated system, it is possible to verify that they are the same up to the substitution $H^{(1)}=H^{(1)}_S$ and the renormalization of $\epsilon$ by the constant factor $\sum_{p} b_p\bra{p} H^{(1)}_E\ket{p}$. Therefore, the robustness condition $\mathcal{R}=0$ in Eq.~\eqref{R} ensures that $\rho^{(1)}_{q,S}(T)=0$.\\

We now consider the final second-order density matrix $\rho_q^{(2)}(T)$, as in Eq.~\eqref{rho_3_bis}, and trace out the environment degrees of freedom to obtain the evolution of the system only
\begin{align}
    \rho^{(2)}_{q,S}(T) &=  \text{Tr}_E\left\{ \rho^{(2)}_{q}(T) \right\}\\
    \nonumber
    =& - \sum_{p,p'} b_p \int_0^T dt \int_0^T ds \bra{p'} U^{(0)}(t,T)H^{(1)} U^{(0)}(0,t)\cdot \\
    & \cdot\ket{q,p}\bra{q,p} U^{(0)}(s,0)H^{(1)}U^{(0)}(T,s)\ket{p'}\\
    \nonumber
    = & - \sum_{p,p'} b_p \int_0^T dt \int_0^T dsU^{(0)}_S(t,T)H^{(1)}_SU^{(0)}_S(0,t)\cdot\\
    \nonumber
    &\cdot \ket{q}\bra{q} U^{(0)}_S(s,0)H^{(1)}_SU^{(0)}_S(T,s) \cdot \\
    \label{expression_rho2}
    &\cdot\left|\bra{p'}H^{(1)}_E\ket{p}\right|^2 e^{-i(t-s)(E_p-E_{p'})},
\end{align}
where the fact that $U_E^{(0)}(s,t)\ket{p} = \exp[-iE_p(t-s)]\ket{p}$ is used, since $\ket{p}$ are eigenstates of $H_E^{(0)}$.\\
In the integral in Eq.~\eqref{expression_rho2}, the exponential term $\exp[-i(t-s)(E_p - E_{p'})]$ oscillates with a frequency $E_p - E_{p'}$ over a time interval $T$, which is typically determined by the energy scale of the system Hamiltonian $H_S^{(0)}$. Assuming that the energy scale of the environment $H_E^{(1)}$ is much larger than that of the system Hamiltonian $H_S^{(0)}$ -- i.e., that $T$ is much longer compared to the time scale set by the frequency $E_p - E_{p'}$ -- one can use a rotating-wave-like approximation to argue that the terms $\bra{p'} H_E^{(1)} \ket{p}$ that contribute significantly are those that are resonant, meaning $\ket{p}$ and $\ket{p'}$ have the same energy, i.e., $E_p = E_{p'}$. This assumption constitutes the adiabatic approximation: the time scale over which the environment evolves is small compered to the time scale over which the state of the system varies appreciably. The expression in Eq.~\eqref{expression_rho2} becomes 
\begin{align}
    \nonumber
    & \rho^{(2)}_{q,S}(T) = - \int_0^T dt \int_0^T dsU^{(0)}_S(t,T)H^{(1)}_SU^{(0)}_S(0,t)\ket{q}\bra{q}\cdot \\
    \label{expression2_rho2}
    & \cdot U^{(0)}_S(s,0)H^{(1)}_SU^{(0)}_S(T,s) \left(\sum_{p,p'}^{E_p=E_{p'}} b_p \left|\bra{p'}H^{(1)}_E\ket{p}\right|^2 \right).
\end{align}
Comparing the expression for the second-order reduced density matrix $\rho_{q,S}^{(2)}$ in Eq.~\eqref{expression2_rho2} with the one in Eq.~\eqref{rho_2_bis} for the second-order density matrix of an isolated system, it is possible to verify that they are the same, up to the substitution $H^{(1)}=H^{(1)}_S$, and the renormalization of $\epsilon^2$ by the constant factor $\sum_{p,p'}^{E_p=E_{p'}} b_p\left|\bra{p'}H^{(1)}_E\ket{p}\right|^2$. Therefore, the robustness condition $\mathcal{R}=0$ in Eq.~\eqref{R} ensures that $\rho^{(2)}_{q,S}(T)=0$.\\
In Sec.~\ref{sect:physical_realizations}, a physical realization is discussed in which adiabatic approximation holds: polar molecules trapped in optical tweezers and interacting via dipole-dipole interactions, where the gate duration $T$ is set by the inverse of the dipole-dipole interaction strength $J$, and the energy scale of the environment is given by the trapping tweezers frequency $\omega$.\\

The results of this section can be summarized by saying that: in the presence of a system $S$ coupled to an environment $E$, as in Eq.~\eqref{H_environment}, and under the  adiabatic approximation -- i.e. a time scale of the environment that is small compared to the time scale of the system -- a quantum operation can be made robust using the criterion $\mathcal{R}=0$ in Eq.~\eqref{R}, tracing out the environment $E$, and using as first-order Hamiltonian the noise operator $H_S^{(1)}$ that acts on the system only.\\
As far as we known this analytical result is new in literature, however, in Ref.~\cite{Giudici_2024} it was numerically shown that, in the context of CZ gates between Rydberg atoms, pulses optimized with respect to quasi-static noise fluctuations preserve their robustness also in presence of a coupling to motion space. It should be pointed out, however, that in that work the dominant source of noise was an off-resonant term, because of the energy scale of the system. A similar numerical confirmation is shown in Fig.~\hyperref[fig:motion_omega]{\ref{fig:motion_omega}(b)} for the molecular iSWAP gate studied in this work.

\section{\label{section_2}The iSWAP gate}

In this section, we apply the general tools discussed above to the specific case of an exchange Hamiltonian implementing an iSWAP gate [Sec.~\ref{sec:exchange_Hamiltonian_iSWAP}]. In Sec.~\ref{sec:robust?_no_sorry}, we show that the criteria \ref{criterion_i} - \ref{criterion_iii} of the previous section rule out the possibility of achieving a robust quantum gate using global control only. In Sec.~\ref{sect:solution}, we propose a protocol to work around this fact by adding local controls. In Sec.~\ref{sec:physical_interpretation}, we provide a physical explanation of why local controls are required.

\subsection{\label{sec:exchange_Hamiltonian_iSWAP}The exchange Hamiltonian and the iSWAP gate}

We consider the following two-qubit interaction Hamiltonian of the exchange type
\begin{align}
    \label{H_0}
    H_{0}(J) & = J\left(\sigma_+^{(1)} \sigma_-^{(2)} + \sigma_-^{(1)} \sigma_+^{(2)}\right)\\
    & = J
    \begin{pmatrix}
    0 & 0 & 0 & 0 \\
    0 & 0 & 1 & 0 \\
    0 & 1 & 0 & 0 \\
    0 & 0 & 0 & 0
    \end{pmatrix}
    ,
\end{align}
where $J$ is the interaction strength, $\sigma_{\pm}^{(j)} = (\sigma_x^{(j)} \mp i \sigma_y^{(j)})/2$ are the rising and lowering Pauli operators of the $j$-th qubit, with $j=1,2$ and the matrix is written in the two-qubit computational basis $Q =\{\ket{00},\ket{01}, \ket{10},\ket{11}\}$. The effect of the exchange-type Hamiltonian in Eq.~\eqref{H_0} is to transfer an excitation from one qubit to another, resulting in a coupling between the states $\ket{01}$ and $\ket{10}$.\\

The exchange interaction Eq.~\eqref{H_0} can create entangled states with associated evolution operator  
\begin{align}
\label{U_hop}
    U_{0}(t) = & e^{-iH_{0}t}\\
    = &
    \begin{pmatrix}
    1 & 0 & 0 & 0 \\
    0 & \cos(Jt) & -i\sin(Jt) & 0 \\
    0 & -i\sin(Jt) & \cos(Jt) & 0 \\
    0 & 0 & 0 & 1
    \end{pmatrix}.
\end{align}
By waiting for a time $\bar{T}=\pi/(2J)$, the unitary operator $U_{0}(\bar{T})$ implements the iSWAP gate, an entangling gate that, along with single-qubit rotations, forms a universal set for quantum computation \cite{Ni_2018}. In many practical scenarios, such as optically trapped polar molecules or neutral atoms, the interaction strength $J$ -- and consequently the characteristic time $\bar{T}$ -- are subject to experimental uncertainties, which limit the fidelity of the entangling operation. The main objective of this work is to propose a general scheme to reduce sensitivity to uncertainties in $J$ in the realization of a iSWAP gate.

\subsection{\label{sec:robust?_no_sorry}Impossibility of making the iSWAP gate robust with global control pulses only}

\subsubsection{Dynamical decoupling pulses}

The problem of robustness cannot be solved simply by applying known dynamical decoupling schemes. The classical global $X^{\otimes2}$, $Y^{\otimes2}$ and $Z^{\otimes2}$ pulses of spin-echo and XY8 sequences \cite{Viola_1998, Viola_2003, Souza_2012, Arenz_2018} commute with the interaction Hamiltonian $H_0$ \eqref{H_0}, making them ineffective. This is a well-known result in the literature, consistent with criterion \ref{criterion_i} from Sec.~\ref{sect:criteria} above. Additional strategies to remove unwanted interactions using more elaborate global dynamical decoupling sequences that symmetrize the interaction, e.g. WAHUHA \cite{Waugh_1968, Cohen_2021} have been proposed. However, these schemes remove the effect of the interaction completely, and thus the entanglement of the final state, which is not purpose of this work.

\subsubsection{Generic global pulses}

On the basis of the successes of optimal quantum control techniques based on an external global drive of the qubits (e.g. a laser irradiating both Rydberg atoms on a resonant ground-Rydberg transition for neutral atoms in Ref. \cite{Jandura_2023}), one may attempt to realize a robust quantum gate by using a global control field, $\Omega(t) = |\Omega(t)|\exp(i\phi(t))$,  e.g. a field irradiating identically both qubits, with control Hamiltonian of the form
\begin{align}
    \nonumber
    H_c^G(\Omega(t))= & |\Omega(t)| \Big[\cos\phi(t)(\sigma_x^{(1)}+\sigma_x^{(2)}) +\\
    \label{H_c}
    & +\sin\phi(t)(\sigma_y^{(1)}+\sigma_y^{(2)})\Big],
\end{align}
where $|\Omega(t)|$ is the amplitude of the (Rabi) frequency of the field coupling resonantly the $\ket{0} \leftrightarrow \ket{1}$ transition, $\phi(t)$ the phase, and $\sigma_{x,y}^{(j)}$ the Pauli $x$ and $y$ matrices acting on the qubit $j$. The total Hamiltonian then reads $H(J,\Omega(t)) = H_0(J) + H_c^G(\Omega(t))$.\\

The quantum optimal control problem in Eq.~\eqref{problem0} is here rephrased as
\begin{equation}
    \label{problem_specific}
    \mathcal{T} \exp\left\{-i \int_0^{T} dt H(J+\Delta J, \Omega(t))\right\} = U_0(\bar{T}) + o(\Delta J^2),
\end{equation}
with $\bar{T}\neq T$, so that the robust pulse in principle can be longer than $\bar{T}=\pi/(2J)$.
With reference to Eq.~\eqref{H_decomposition} here $H^{(0)}(t) = J(\sigma_+^{(1)} \sigma_-^{(2)} + \sigma_-^{(1)} \sigma_+^{(2)}) + H_c^G(\Omega(t))$ is the zeroth-order Hamiltonian in $J$, and the first-order Hamiltonian in $J$ is $H^{(1)} = \sigma_+^{(1)} \sigma_-^{(2)} + \sigma_-^{(1)} \sigma_+^{(2)} $.\\

The operator $H^{(1)}$ can be written as $H^{(1)} = S \mathbb{P}_1$, where $S$ is the SWAP operator and $\mathbb{P}_1$ is the projector on the states $\{\ket{01},\ket{10}\}$
\begin{equation}
\label{SWAP}
    S =
    \begin{pmatrix}
        1 & 0 & 0 & 0 \\
        0 & 0 & 1 & 0 \\
        0 & 1 & 0 & 0 \\
        0 & 0 & 0 & 1
    \end{pmatrix}, \quad
    \mathbb{P}_1 =
    \begin{pmatrix}
        0 & 0 & 0 & 0 \\
        0 & 1 & 0 & 0 \\
        0 & 0 & 1 & 0 \\
        0 & 0 & 0 & 0
    \end{pmatrix}.
\end{equation}
For a global control $H_c^G$, as in Eq.~\eqref{H_c}, acting symmetrically on both qubits and thus invariant under SWAP action, the SWAP operator in Eq.~\eqref{SWAP} commutes with the $H^{(0)}$ Hamiltonian. Moreover the operator $\mathbb{P}_1$ is positive semi-definite. Therefore, the criterion \ref{criterion_iii} is satisfied, making it impossible to render the gate robust against perturbations in the interaction strength $J$ using the global control Hamiltonian $H_c^G$ in Eq.~\eqref{H_c} or any other global control.\\
The reason that makes this problem different from e.g. the CZ gates with a blockade interaction \cite{Jandura_2023}, is that the Hamiltonian at the origin of the noise and the one that implements the desired evolution -- and creates entanglement -- are the same.\\

\subsection{\label{sect:solution}Robustness by local control}

As seen in the Sec.~\ref{sec:robust?_no_sorry} above, the key feature that makes it impossible to create robust pulses by adding an external global control is the fact that $H^{(1)}$ can be decomposed in two matrices, one ($\mathbb{P}_1$) that is positive semi-definite, the other ($S$) that commutes with $H^{(0)}$. This argument breaks down if $H^{(0)}$ is modified in such a way that it no longer commutes with $S$, i.e. by breaking the exchange symmetry between the two molecules.\\
In the zeroth-order Hamiltonian $H^{(0)}$, there is the control Hamiltonian $H_c$, which may be modified to act non-symmetrically on the two qubits. Several approaches can be used to achieve this, depending on the physical setup and the single-site addressability of the experimental system. For instance, assuming full single-site addressability, the control Hamiltonian has two different Rabi frequencies $|\Omega_j|$ and phases $\phi_j$ for each qubit $j=1,2$. In this case the control Hamiltonian reads
\begin{align}
    \nonumber
    H_c^A(\Omega_{1,2}(t))= &\sum_{j=1}^2|\Omega_j(t)| \Big[\cos\phi_j(t)\sigma_x^{(j)} +\\
    \label{H_c1}
    & + \sin\phi_j(t) \sigma_y^{(j)} \Big]. 
\end{align}
In the case of qubits encoded in the rotational hyperfine states of molecules, as discussed in Sec.~\ref{sect:physical_realizations}, the local drive $\Omega_{j}(t)$ that appears inside the control $H_c^A(t)$ \eqref{H_c1} is a microwave field. Realizing such local microwave controller is experimentally challenging, but it is also not necessary. To break the exchange symmetry, it is, in fact, sufficient to use a constant local detuning $\delta \sigma_z^{(2)}$, combined with an optimal global drive $\Omega(t)$. The local detuning can be implemented relatively easily either by adjusting the intensity of tweezers or by producing light shifts with another light beam. In this way the control Hamiltonian reads
\begin{align}
    \nonumber
    H_c^B(\Omega(t))= & |\Omega(t)| \Big[\cos\phi(t)(\sigma_x^{(1)}+\sigma_x^{(2)}) +\\
    \label{H_c2}
    & +\sin\phi(t)(\sigma_y^{(1)}+\sigma_y^{(2)})\Big] + \delta\sigma_z^{(2)}.
\end{align}
In both cases above the analytic arguments \ref{criterion_i}, \ref{criterion_ii} and \ref{criterion_iii} do not apply and thus it should be possible to design robust pulses. This is achieved using quantum optimal control techniques [in Sec.~\ref{section_3} below].\\

It is worth noting that a related result has been derived in the context of controllability of spin-$1/2$ networks, where the need for local controls also arises. It has been shown that in a chain of spins with Heisenberg-type interactions, complete operator controllability -- i.e., the ability to implement arbitrary unitary transformations on the system's Hilbert space -- cannot be achieved using only global controls. Instead, the latter requires acting on a single qubit of the chain with two local, non-commuting control operators \cite{Heule_2010, Wang_2016}. A similar result holds for a spin chain with Ising interactions \cite{Albertini_2018, Stojanovic_2022, Stojanovic_2023}.
These results pertain to the ability of controls to generate arbitrary unitary evolutions, without concern for robustness, i.e., not distinguishing between a robust iSWAP gate and a non-robust one. The criteria \ref{criterion_i}-\ref{criterion_iii} discussed in this work instead focus on the ability of controls to generate \textit{robust} unitary evolutions. Therefore, even if the realization of a non-robust iSWAP gate does not require local controls, the realization of a robust iSWAP instead requires local controls.

\subsection{\label{sec:physical_interpretation}Physical interpretation}

To provide a physical interpretation of the necessity for a local control Hamiltonian, consider the following basis $Q' = \{\ket{11}, \ket{\Psi_+}, \ket{00}, \ket{\Psi_-}\}$, where $\ket{\Psi_\pm} = (\ket{01} \pm \ket{10})/\sqrt{2}$. In this basis, the exchange Hamiltonian $H_0$ \eqref{H_0} is diagonal \cite{Hughes_2020}
\begin{equation}
    \label{H_0'}
    H_0(J) =
    \begin{pmatrix}
        0 & 0 & 0 & 0\\
        0 & J & 0 & 0\\
        0 & 0 & 0 & 0\\
        0 & 0 & 0 & -J
    \end{pmatrix}.
\end{equation}
The Hilbert space of two qubits can now be decomposed into the direct sum of two disjoint subspaces, according to how states behave under the action of the SWAP operator. The symmetric subspace is spanned by $\{\ket{11}, \ket{\Psi_+}, \ket{00}\}$, and it is invariant under the exchange of the two qubits. The antisymmetric subspace is spanned by $\{\ket{\Psi_-}\}$, and it acquires a factor of $-1$ under the action of a SWAP. As shown by the Hamiltonian in Eq.~\eqref{H_0'}, the exchange interaction does not mix the two subspaces.\\
                                                                           
Within the symmetric subspace, the effects of the noise $\Delta J$ on the interaction strength $J$ can not be canceled as the symmetric part of the Hamiltonian in Eq.~\eqref{H_0'} is positive semi-definite. Indeed the states $\ket{00}$ and $\ket{11}$ are unaffected by the interaction (and thus by its noise), while the state $\ket{\Psi_+}$ acquires an unwanted phase proportional to $\Delta J$ and the time spent in the $\ket{\Psi_+}$ state. To eliminate this extra phase, the only option is to access the antisymmetric subspace, where a noisy interaction results in an extra phase proportional to $-\Delta J$ times the time spent in the $\ket{\Psi_-}$ state. This effect can be used to cancel the overall effect of the noise.

\begin{figure*}[ht]
    \centering
   \includegraphics[width=0.98\linewidth]{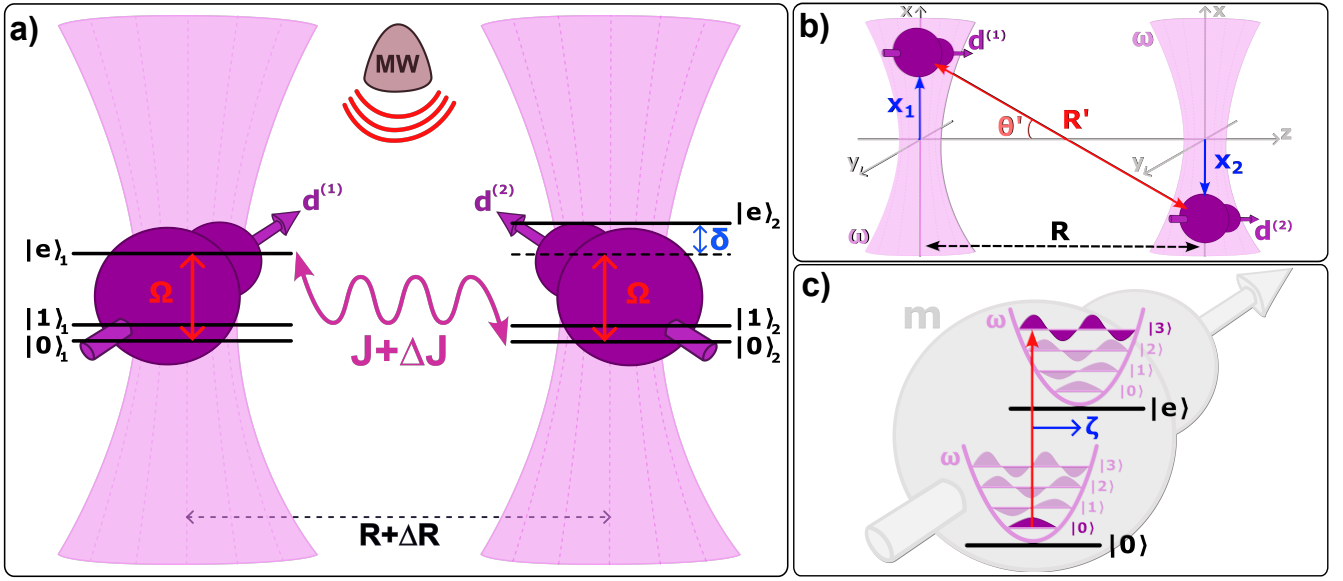}
    \caption{\textbf{a)} Schematic representation of two polar molecules trapped in optical tweezers, at a distance $R$. The logical states $\{\ket{0}_j,\ket{1}_j\}$ ($j=1,2$) are encoded in hyperfine levels of the rotational ground state, while the rotational excited state $\ket{e}_1$ is strongly coupled to $\ket{0}_2$ via dipole-dipole interaction of strength $J$, which depends on $R$ (clearly the same holds for $\ket{e}_2$ and $\ket{0}_1$). Perturbations $\Delta R$ in the intermolecular distance $R$ generate perturbations $\Delta J$ on the coupling $J$. Letting the two molecules interacting for a time $\bar{T}=\pi/(2J)$ one realizes the entangling iSWAP gate, and by adding local controls, like the microwave controllers $\Omega_1(t)$ and $\Omega_2(t)$ [see Eq.~\eqref{H_c2}], one can make the gate robust to the noise $\Delta J$. \textbf{b)} Schematic representation of the effect of axial motion on the intermolecular distance $R$ and the angle $\theta$. Thermal motion induces a variation in the intermolecular distance, in particular $R'=\sqrt{R^2+(x_1-x_2)^2}$, where $x_j$ is the position of the $j$-th molecule, and in the angle $\theta$, such that $\cos\theta'=R/R'$. Consequently, also the coupling strength $J'=d^2(1-3\cos^2\theta')/R'^3$ is perturbed.  \textbf{c)} Schematic representation of the motion-rotation coupling. In a molecule of mass $m$ trapped in an optical trap of frequency $\omega$ the motion-rotation coupling is originated by optical aberrations that cause a displacement $\zeta$ (blue arrow) of the trapping potential of the first excited rotational state $\ket{e}$, with respect to the rotational ground state $\ket{0}$ (in black). Rotational transitions $\ket{0}\leftrightarrow\ket{e}$ (red arrow) are much faster than molecular motion in the optical traps, therefore the transition that has larger probability to happen is the one that maximizes the overlap between the initial and final harmonic eigenfunctions (in violet), thus the molecule undergoes not only a rotational transition, but also a motional transition.}
    \label{fig:molecoles}
\end{figure*}

\section{\label{sect:physical_realizations}Polar molecules as qubits}

In Sec.~\ref{sec:polar_molecules}, we present a physical realization of the exchange Hamiltonian for ultracold polar molecules prepared in their electronic and vibrational ground state, where the dipole-dipole interaction couples different rotational manifolds. In Sec.~\ref{sec:mot-rot_coupling}, we discuss the main relevant sources of noise in this setup, in particular the motion in the axial direction and the motion-rotation coupling, that determine variations in the dipole-dipole interaction strength, and estimate their magnitude. We also describe in which experimental parameter regime the adopted adiabatic approximation is valid. This system represents a possible interesting application of the robust protocols illustrated in Sec.~\ref{sect:results} below.

\subsection{\label{sec:polar_molecules}Polar molecules}

One important physical realization of the exchange interaction in Eq.~\eqref{U_hop} is with cold polar molecules trapped in optical tweezers [see Fig. \hyperref[fig:molecoles]{\ref{fig:molecoles}(a)}], with qubit logical states $\{\ket{0},\ket{1}\}$ encoded in hyperfine sublevels of the electronic, vibrational and rotational ground state manifold and interacting via dipole-dipole interaction (see below) \cite{Ni_2018, Cornish_2024, Picard_2024_arXiv}. \\

As an example, we consider NaCs polar molecules with qubit states encoded into the ground hyperfine states $\ket{0} = \ket{N=0, m_N=0, m_{\text{Na}}=3/2, m_{\text{Cs}}=5/2}$ and $\ket{1} = \ket{N=0, m_N=0, m_{\text{Na}}=-1/2, m_{\text{Cs}}=5/2}$ \cite{Ni_2018, Picard_2024_arXiv}. Here $N$ and $m_N$  are the quantum numbers for the rotational angular momentum and its projection onto the quantization axis, respectively, whereas $m_{\text{Na}}$ and $m_{\text{Cs}}$ are the projections of the nuclear spin of Sodium and Cesium onto the quantization axis, respectively. The chosen qubit states $\ket{0}$ and $\ket{1}$ have long coherence times of the order of 80 ms \cite{Picard_2024_arXiv}.\\

The dipole-dipole interaction between two polar molecules reads \cite{Ni_2018, Hughes_2020}
\begin{equation}
    \label{H_dd}
    H_{dd} = \frac{1}{4\pi\epsilon_0R^3} \left( \mathbf{d}^{(1)}\cdot\mathbf{d}^{(2)}-3(\mathbf{d}^{(1)}\cdot\mathbf{\hat{R}})\otimes(\mathbf{d}^{(2)}\cdot\mathbf{\hat{R}}) \right),
\end{equation}
where $R$ is the molecular separation, $\mathbf{\hat{R}}$ denotes the unit vector pointing from one qubit to the other, and $\mathbf{d}^{(j)}$ is the electric dipole moment operator for the molecule $j$. 
The dipole-dipole interaction Hamiltonian $H_{dd}$ of Eq.~\eqref{H_dd} implements the exchange interaction Hamiltonian $H_0$ of Eq.~\eqref{H_0}. Assuming parallel dipoles, the coupling strength is $J=d^2(1-3\cos^2\theta)/R^3$ \cite{Picard_2024_arXiv}, where $\theta$ is the angle between the transition dipole moment $\mathbf{d}^{(j)}$ and the intermolecular axis $\mathbf{\hat{R}}$, and $d$ is the amplitude of the transition dipole moment $\mathbf{d}^{(j)}=d \sigma_x^{(j)}$, with $\sigma_x^{(j)}$ the Pauli $x$ matrix acting on the two interacting states of the qubit $j$.\\ 

The chosen qubit states $\ket{0}$ and $\ket{1}$ have long coherence times, but zero common transition dipole moment $d_{10}=\bra{0}\mathbf{d}\ket{1}=0$, implying no direct dipole-dipole interactions for qubits prepared in these states.
The dipole-dipole interaction between the qubits can be activated (``toggled") by temporarily transferring the population of the $\ket{1}$ state to a given state $\ket{e}$ in the first rotational manifold ($N=1$) which has a large transition dipole moment $\mathbf{d}_{e0}$ with the ground state $\ket{0}$, using a microwave Rabi $\pi$-pulse. Following the experiment of Ref. \cite{Picard_2024_arXiv}, here we choose $\ket{e}=\ket{N=1, m_N=1, m_{\text{Na}}=3/2, m_{\text{Cs}}=5/2}$, that has $d_{e0}=4.6/\sqrt{3} \text{ Debye}$.\\

By properly timing the time spent in the excited state $\ket{e}$, the two qubits can now exchange an excitation energy (corresponding to rotational energy) via the dipole-dipole interaction $H_{dd}$ in Eq.~\eqref{H_dd} via the $\ket{0}\leftrightarrow\ket{e}$ transition, thus realizing the exchange Hamiltonian $H_0$ in Eq.~\eqref{H_0}. The population of the $\ket{e}$ state is then transferred back to the ground state $\ket{1}$ \cite{Picard_2024_arXiv}, finally realizing the desired entangling gate between the $\ket{0}$ and $\ket{1}$ states. This toggle interaction involving the excited rotational state $\ket{e}$ thus allows one to ``switch" the dipole-dipole interaction on and off and thus to perform the entangling gate when needed.\\

Electric dipole moments $d$ of diatomic polar molecules are usually in the range $0.5-5 \text{ Debye}$, whereas the inter-molecular spacing $R$ is typically in the range $0.5$–$3$ $\mu\text{m}$ \cite{Cornish_2024}. With a dipole moment of $d = 4.6/\sqrt{3} \text{ Debye}$ and inter-molecular spacing of $R=1.9$ $\mu\text{m}$, the interaction strength results to be $J = 2\pi \times 0.715 \text{ kHz}$ \cite{Picard_2024_arXiv}. The dominant source of noise for this type of gate is its sensitivity to uncertainties in the dipole-dipole interaction strength $J$. These uncertainties arise from static deviations due to imprecisions in the relative positioning of the confining potential, as well as from the relative motion of the molecules within the optical traps. In Sec.~\ref{sec:mot-rot_coupling} below, we analyze these noise sources and their effects in detail.

\subsection{\label{sec:mot-rot_coupling}Noise in the interaction strength}

In this section, we discuss the main relevant source of noise with optically trapped molecules, in particular the thermal motion and the motion-rotation coupling, showing that the former is the dominant source of decoherence. We also precisely describe in which experimental parameter regime the quasi-static noise approximation, adopted in the previous sections, is valid.

\subsubsection{Axial motion}
\label{sec:axial_mot}

Experiments show that the dominant dephasing mechanism within an entangling operation is variation of the interaction strength $J$ \cite{Holland_2023, Bao_2023, Cornish_2024, Picard_2024_arXiv, Ruttley_2025}, due to, e.g. relative motion of the molecules during the gate, in particular along the axial direction of the traps. The latter effect causes fluctuations in both the molecular separation $R$ and the dipole orientation $\theta$.\\
The three-dimensional trapping potential of the optical tweezers can be modeled as three harmonic oscillators of frequency $\omega_\alpha$, with $\alpha=x,y,z$. The motional degrees of freedom along the direction $\alpha$ of each molecule $j$ are governed by the following Hamiltonian
\begin{equation}
\label{H_mot}
    H_{\alpha,j}^{\text{mot}} = \omega_\alpha\left(a_{\alpha,j}^\dag a_{\alpha,j} + \frac{1}{2}\right),
\end{equation}
where $a_{\alpha,j}$ and $a_{\alpha,j}^\dag$ are the annihilation and creation operators, respectively, of the harmonic motional modes in the direction $\alpha$ for the molecule $j$.  The single-molecule Hilbert space now is the tensor product of the rotation Hilbert space $\mathcal{H}^{\text{rot}}_j$, spanned by the computational basis $\{\ket{0}_j, \ket{e}_j\}$, and the motion Hilbert spaces $\mathcal{H}^{\text{mot}}_{\alpha,j}$, generated by the action of the creation operators $a_{\alpha,j}^\dag$ on the vacuum state.\\
The position of the $j$-th molecule with respect to the corresponding trap center is represented by the displacement vector $\vec{r}_j=(x_j, y_j,z_j)$, where the $z$ and $y$ axes denote the radial directions of the tweezers, while $x$ the axial one. In particular, $z$ designates the interatomic direction, as depicted in Fig. \hyperref[fig:molecoles]{\ref{fig:molecoles}(b)}.  If $\vec{r}_j=0$, the intermolecular distance is simply $\vec{R}= (0,0,R)$, but in the presence of thermal motion -- and even at zero temperature due to the finite width of the motional ground-state wavefunction --  the displacement vectors satisfy $\vec{r}_j\neq0$ and the perturbed intermolecular distance becomes $\vec{R}'= \vec{R}+\vec{r}_1-\vec{r}_2$, with length $R'$
\begin{equation}
    \label{R_perturbed}
    R'=\sqrt{(x_1-x_2)^2+(y_1-y_2)^2+(R+z_1-z_2)^2}.
\end{equation}
Typically, the trapping potentials along the radial directions, $\omega_{y}$ and $\omega_z$, are one order of magnitude larger than the axial one, $\omega_x$. Therefore, the dominant motion occurs along the axial direction, while the radial contributions are small corrections. From now on, the axial trapping frequency is simply denoted as $\omega$. Moreover, the characteristic harmonic length, $1/\sqrt{2m\omega}$, with $m$ the mass of the molecule, is on the order of tens of nanometers, whereas the intermolecular distance is on the order of micrometers. Thus, we can expand the perturbed intermolecular distance $R'$ in Eq.~\eqref{R_perturbed} in powers of $(x_1 - x_2)/R$. In particular, recalling that the coupling strength $J \propto R^{-3}$, the inverse of the cube of the perturbed intermolecular distance $R'$ in Eq.~\eqref{R_perturbed}, up to second order in $(x_1-x_2)/R$, reads 
\begin{equation}
    \label{1/R^3}
    \frac{1}{R'^3}=\frac{1}{R^3}\left[1-\frac{3}{2}\left(\frac{x_1-x_2}{R}\right)^2\right].
\end{equation}
The interaction strength $J$ also depends on the angle $\theta$ between the dipole moments $\mathbf{d}^{(j)}$ and the intermolecular axis $\hat{z}$. In the unperturbed case, this angle is simply $\theta = 0$. However, in the presence of motion, it becomes $\cos\theta' = \hat{z} \cdot \vec{R}'/R' = R/R'$, where $\hat{z}$ is a unit vector along the $z$ axis.
Recalling that the coupling strength scales as $J \propto 1 - 3\cos^2\theta$, we can expand the angular part of $J$ up to second order in $(x_1 - x_2)/R$, yielding
\begin{equation}
    \label{theta'}
   1-3\cos^2\theta' = -2 + 3\left(\frac{x_1-x_2}{R}\right)^2.
\end{equation}
Combining the effects of motion on both intermolecular distance $R$ and angle $\theta$, Eqs. \eqref{1/R^3} and \eqref{theta'} respectively, it is possible to estimate the perturbed coupling constant $J'=d^2(1-3\cos^2\theta')/R'^3$ as
\begin{align}
    J'& = \frac{d^2}{R^3} \left[ -2 + 6 \left(\frac{x_1-x_2}{R}\right)^2\right]\\
    \label{J'}
    & = J \left[ 1 - 3 \left(\frac{x_1-x_2}{R}\right)^2\right],
\end{align}
where in the last line the fact that in this frame $J=-2d^2/R^3$ has been used.\\
In the presence of motion, the dipole-dipole interaction Hamiltonian $H_0$ in Eq.~\eqref{H_0}, using the result in Eq.~\eqref{J'}, then becomes
\begin{equation}
    \label{H_0_delta_R}
    H_0 = \frac{J}{2} \left[1- 3 \left(\frac{x_1-x_2}{R}\right)^2\right]\left(\sigma_x^{(1)} \sigma_x^{(2)} + \sigma_y^{(1)} \sigma_y^{(2)}\right).
\end{equation}
One can then estimate the magnitude of the effect of motion as
\begin{equation}
    \label{Delta_J_R}
    \frac{\Delta J_{\text{mot}}}{J} = \frac{3\sigma_{(x_1-x_2)^2}}{R^2}.
\end{equation}
Here, $\sigma_{(x_1 - x_2)^2}$ denotes the standard deviation of the squared position difference operator $(x_1 - x_2)^2$. It can be computed by assuming that both molecules $j$ are initially in a thermal state, $\rho_j^{\text{th}} \propto \sum_n \exp\left[- H_j^{\text{mot}} / (k_B T_{\text{temp}})\right]\ket{n}\bra{n}$, where $H_j^{\text{mot}}$ is the axial harmonic oscillator Hamiltonian defined in Eq.~\eqref{H_mot}, $k_B$ the Boltzmann constant, and $T_{\text{temp}}$ the temperature.
The expected value of the squared position operator $x_j^2$ over the thermal state $\rho_j^{\text{th}}$ is
\begin{equation}
    E_{\text{th}}[x_j^2] = \Tr\left\{\rho_j^{\text{th}}x_j^2\right\} = \frac{1}{2m\omega} \coth\left(\frac{\omega}{2k_BT_{\text{temp}}}\right),
\end{equation}
and analogously
\begin{equation}
    E_{\text{th}}[x_j^4] = \Tr\left\{\rho_j^{\text{th}}x_j^4\right\} = \frac{3}{(2m\omega)^2} \coth^2\left(\frac{\omega}{2k_BT_{\text{temp}}}\right).
\end{equation}
For a generic operator $A$, the standard deviation is given by $\sigma_A = \sqrt{E_{\text{th}}[A^2] - E_{\text{th}}[A]^2}$. By noting that $E_{\text{th}}[x_j] = 0$ and that $\sigma_{(x_1 - x_2)^2} = \sqrt{2}  \sigma_{x_1 - x_2}^2$, one obtains an estimate of the noise parameter $\Delta J_{\text{mot}} / J$ as expressed in Eq.~\eqref{Delta_J_R} as
\begin{equation}
    \label{Delta_J_mot}
    \frac{\Delta J_{\text{mot}}}{J}=6 \sqrt{2} \frac{1/(2m\omega)}{R^2} \coth\left(\frac{\omega}{2k_BT_{\text{temp}}}\right),
\end{equation}
thus the effect of axial motion depends on the ratio between the harmonic oscillator length $\sqrt{1/(2m\omega)}$ and the intermolecular distance $R$, but also on the ratio between the confining potential energy $\omega$ and the thermal energy $k_BT_{\text{temp}}$. In the experiment in Ref.~\cite{Picard_2024_arXiv}, the values of these parameters were estimated, as $\sqrt{1/(2m\omega)}=80\text{ nm}$, $R=1.9$ $\mu$m and $\omega/(k_BT_{\text{temp}})=0.42$. Therefore, $\Delta J_\text{mot}/J\approx7.3\times10^{-2}$. It is possible to directly verify that the neglected terms -- such as radial motion and higher-order contributions -- are small corrections. It is worth noting that the first-order radial contributions are off-resonant, as we discuss in the section below on first-order term of motion-rotation coupling.

\subsubsection{Motion-rotation coupling}

Another effect that can potentially affect the coherence of the exchange interaction is the coupling of the rotational states to the harmonic motion within the axial trap potential \cite{Picard_2024_arXiv}. At the origin of the coupling are optical aberrations, e.g. astigmatism, that cause a displacement $\zeta$ of the trapping potential center for the first excited rotational state $\ket{e}$, with respect to the rotational ground state $\ket{0}$, as shown in Fig. \hyperref[fig:molecoles]{\ref{fig:molecoles}(b)}. This effect originates is the experiment of Ref.~\cite{Picard_2024_arXiv}, arising from the choice of a "magic" polarization of the optical tweezers, which compensates for the differential light shift \cite{Rosenband_2018}.\\
The Franck-Condon effect is a well-known phenomenon describing how, due to a displacement in the electronic potential energy surfaces, an electronic transition in a molecule can also be accompanied by a change in its vibrational state \cite{Franck_1926, Condon_1926, Condon_1928}. The key idea is that electronic transitions occur so rapidly compared to the nuclear vibrational motions that the nuclei can be considered to remain in the same position during the transition. The probability distribution for the final vibrational states is governed by the overlap between the initial and final vibrational wavefunctions.\\
Analogously, rotational transitions $\ket{0}\leftrightarrow\ket{e}$ typically occur much faster than the timescales of molecular motion in the harmonic optical traps. Therefore, the transition with the highest probability is the one that maximizes the overlap between the initial and final harmonic motion eigenfunctions, leading to excitations in the motional space.\\

A minimal coupling model with a state-dependent displacement of the axial trapping potential for the molecule $j$ is realized by the Holstein Hamiltonian \cite{Mahan_2013, Wellnitz_2022, Picard_2024_arXiv}
\begin{equation}
    \label{H_mot_rot}
    H^{\text{mot-rot}}_j = - \frac{\zeta \omega}{2}(a_j+a_j^\dag) \ket{e}\bra{e}_j,
\end{equation}
where $\zeta$ is the displacement in units of $\sqrt{1/(2m\omega)}$. The interaction Hamiltonian $H_j^{\text{mot-rot}}$ in Eq.~\eqref{H_mot_rot} corresponds to a shift of the harmonic potential energy of the excited state $\ket{e}$ with respect to that of the ground state $\ket{0}$ in the molecule $j$.\\
A similar effect arises also in the presence of an external control fields, when the absorption or the emission of a photon are associated with a momentum kick $\exp[-i\eta(a_j+a_j^\dag)]$ \cite{Giudici_2024}. However, due to the small Lamb-Dicke parameter $\eta$ associated with the microwave transition, the momentum kick here can be neglected.

\subsubsection{Effect of the motion-rotation coupling on the dipole-dipole interaction}

Including the motion-rotation coupling Hamiltonian $H^{\text{mot-rot}}_j$ in Eq.~\eqref{H_mot_rot}, the total Hamiltonian $H_{\text{\text{tot}}}$ for two molecules interacting by dipole-dipole interaction and in the absence of external fields reads
\begin{align}
    \label{H_tot_1}
    H_{\text{\text{tot}}} = & \sum_{i=1}^2 H_j + H_0(J)\\
    \nonumber
    = & \sum_{j=1}^2 \left( H_j^{\text{mot}} + \frac{\Delta}{2}\sigma_z^{(j)} - \frac{\zeta \omega}{2}(a_j+a_j^\dag) \ket{e}\bra{e}_j  \right)+\\
    \label{H_tot_2}
    & +\frac{J}{2}\left(\sigma_x^{(1)} \sigma_x^{(2)} + \sigma_y^{(1)} \sigma_y^{(2)}\right),
\end{align}
where in Eq.~\eqref{H_tot_1} $H_j$ is the single-qubit Hamiltonian for the qubit $j$, and $H_0(J)$ the exchange interaction Hamiltonian from Eq.~\eqref{H_0}. In Eq.~\eqref{H_tot_2}, $H_j^{\text{mot}}$ is the axial harmonic oscillator Hamiltonian in Eq.~\eqref{H_mot}, corresponding to the $j$-th qubit, and $\Delta$ the energy splitting between the states $\ket{0}$ and $\ket{e}$. The energy splitting $\Delta$, the displacement $\zeta$ and the axial trapping frequency $\omega$ are assumed to be the same for both qubits.\\
To investigate the effects of the motion-rotation coupling, we perform a unitary transformation $U_j$ acting on the qubit $j$, with the goal to remove the motion-rotation coupling and obtaining the effective dipole-dipole interaction Hamiltonian in this frame. The chosen unitary is
\begin{equation}
    \label{U_unitary}
    U_j = \exp\left[-\frac{\zeta}{4}\left(a_j-a^\dag_j\right)\left(\mathbb{I}^{(j)}-\sigma_{z}^{(j)}\right)\right],
\end{equation}
where $(\mathbb{I}^{(j)}-\sigma_{z}^{(j)})/2=\ket{e_j}\bra{e_j}$ with $\mathbb{I}^{(j)}$ the identity operator on the rotation Hilbert space of the qubit $j$. Using Campbell-Baker–Hausdorff identity for non-commutative matrices $A$ and $B$ \cite{Sakurai_2020}
\begin{equation}
    e^A B e^{-A} = B + [A,B] + \frac{1}{2!}\left[A,[A,B]\right] + ... 
\end{equation}
one can compute how the operators in the total Hamiltonian $H_{\text{tot}}$ in Eq.~\eqref{H_tot_2} transform under the unitary transformation $U_j$ in Eq.~\eqref{U_unitary}. Namely
\begin{align}
    \label{trans_1}
    U_ja_jU_j^\dag = & a + \frac{\zeta}{4}\left(\mathbb{I}^{(j)}-\sigma_{z}^{(j)}\right),\\
    U_ja_j^\dag U_j^\dag = & a^\dag + \frac{\zeta}{4}\left(\mathbb{I}^{(j)}-\sigma_{z}^{(j)}\right),\\
    U_j\sigma_z^{(j)}U_j^\dag = & \sigma_z^{(j)},\\
    \label{trans_4}
    U_j\sigma_x^{(j)}U_j^\dag = & \cos(2i\Pi_j)\sigma_x^{(j)} - \sin(2i\Pi_j)\sigma_y^{(j)},\\
    \label{trans_5}
    U_j\sigma_y^{(j)}U_i^\dag = & \cos(2i\Pi_j)\sigma_y^{(j)} + \sin(2i\Pi_j)\sigma_x^{(j)},
\end{align}
where $\Pi_j$ is an operator acting on the motional Hilbert space of the qubit $j$ defined as
\begin{equation}
    \Pi_j = -\frac{\zeta}{4}(a_j-a_j^\dag)=-i \frac{\zeta}{\sqrt{8m\omega}}p_j,
\end{equation}
where $p_j=i\sqrt{m\omega/2}(a_j^\dag-a_j)$ is the momentum operator of the axial harmonic trap $j$. Notice that the functions $\cos$ and $\sin$ appearing in Eqs. \eqref{trans_4} and \eqref{trans_5} have to be intended as classical Taylor expansions in powers of the argument operator, e.g. $\cos(2i\Pi_j) =\sum_{n=0}^\infty (2i\Pi_j)^{2n} /((2n)!)$. The displacement $\zeta$ represents the noise parameter and is typically a small quantity, therefore, in the following, only terms up to second order in $\zeta$ are retained, neglecting higher order terms. In this way $\cos(2i\Pi_j)\approx1-(2i\Pi_j)^2/2$ and $\sin(2i\Pi_j)\approx2i\Pi_j-(2i\Pi_j)^3/3!$.\\

Given the transformation rules in Eqs.~\eqref{trans_1}-\eqref{trans_5} and neglecting the terms of order $o(\zeta^3)$ or higher, the total Hamiltonian $H_{\text{tot}}$ in Eq.~\eqref{H_tot_2} under the unitary transformation $U=U_1U_2$ reads
\begin{align}
 \tilde{H} = &  UH_{\text{\text{tot}}}U^\dag\\
 \nonumber
  = & \sum_{j=1}^2 \left( H_j^{\text{mot}} + \frac{\Delta}{2} \sigma_z^{(j)}\right)+\\
    \nonumber
    & + \frac{J}{2} \Bigg[ \left(1-\frac{\zeta^2}{4m\omega}(p_1-p_2)^2\right) \left( \sigma_x^{(1)} \sigma_x^{(2)} + \sigma_y^{(1)} \sigma_y^{(2)} \right)+\\
    \label{H_tot_3}
    & + \frac{\zeta}{\sqrt{2m\omega}}(p_1-p_2) \left( \sigma_x^{(1)} \sigma_y^{(2)} - \sigma_y^{(1)} \sigma_x^{(2)} \right) \Bigg] + o(\zeta^3).
\end{align}
The term at first order in $\zeta$ in Eq.~\eqref{H_tot_3} is proportional to the operator $p_1-p_2$, in which, however, there are no combinations of creation and annihilation operators that preserve the total number of motional excitation in the harmonic traps. Thus this first order term is not resonant and the detuning corresponds to the energy cost of changing a motion level $\omega$, that typically is one order of magnitude larger than the interaction energy scale $J$ [see also Sec.~\ref{subsec:coupling_noise}]. Thus, the term at second order in $\zeta$ in Eq.~\eqref{H_tot_3} represents the most relevant correction to the dipole-dipole interaction strength $J$, since in the operator $(p_1-p_2)^2$ there are combinations of creation and annihilation operators that preserve the total number of motional excitations in the harmonic traps -- e.g. $a^\dag_{1}a_{1}$, $a^\dag_{2}a_{2}$, $a^\dag_{1}a_{2}$ and $a^\dag_{2}a_{1}$ -- thus it is resonant. The magnitude of this resonant noisy term is
\begin{equation}
    \label{Delta_J_zeta}
    \frac{\Delta J_{\text{mot-rot}}}{J} = \frac{\zeta^2}{4m\omega} \sigma_{(p_1-p_2)^2}, 
\end{equation}
where $\sigma_{(p_1-p_2)^2}$ is the standard deviation of the squared of momentum difference operator $(p_1-p_2)^2$. Following the same procedure as above we can derive the expectation value of the squared momentum operator $p_j$ over the thermal state $\rho_j^{\text{th}}$ as
\begin{equation}
    E_{\text{th}}[p_j^2] = \Tr\left\{\rho_j^{\text{th}}p_j^2\right\} = \frac{m\omega}{2} \coth\left(\frac{\omega}{2k_BT_{\text{temp}}}\right).
\end{equation}
and
\begin{equation}
    E_{\text{th}}[p_j^4] = \Tr\left\{\rho_j^{\text{th}}p_j^4\right\} = 3\left(\frac{m\omega}{2}\right)^2 \coth^2\left(\frac{\omega}{2k_BT_{\text{temp}}}\right).
\end{equation}
This yields to estimate the noise parameter $\Delta J_\text{mot-rot}/J$ in Eq.~\eqref{Delta_J_zeta} as
\begin{equation}
    \frac{\Delta J_{\text{mot-rot}}}{J}= \frac{\sqrt{2}}{4} \zeta^2\coth\left(\frac{\omega}{2k_BT_{\text{temp}}}\right).
\end{equation}
Thus the effect of motion-rotation coupling depends on the magnitude of the displacement $\zeta$ and also on the ratio between the confining potential energy $\omega$ and the thermal energy $k_BT_{\text{temp}}$. In the experiment in Ref. \cite{Picard_2024_arXiv}, the values of these parameters are estimated, resulting in $\zeta=0.062$ and $\omega/(k_BT_{\text{temp}})=0.42$. Therefore $\Delta J_\text{mot-rot}/J\approx6.6\times10^{-3}$,  one order of magnitude smaller than the effects of axial motion discussed above. This confirms that the dominant source of decoherence is motion along axial axis.

\subsubsection{adiabatic approximation}

In Sec.~\ref{sec:noise-coupling_env}, it has been shown that pulses robust to quasi-static fluctuations of Hamiltonian parameters preserve their robustness even in the presence of coupling to a noisy environment, e.g. thermal bath. This result holds within the adiabatic approximation, that is, when the energy scale of the system is much smaller than that of the environment. In this section, we discussed the experimental parameter regime in which this approximation is valid.\\
In the physical realization described above, the system consists of two trapped polar molecules, and the corresponding free Hamiltonian $H_S^{(0)}$ is given by the dipolar interaction in Eq.~\eqref{H_0}, i.e., $H_S^{(0)} = J(\sigma_+^{(1)} \sigma_-^{(2)} + \sigma_-^{(1)} \sigma_+^{(2)})$. The noisy environment to which the system is coupled is represented by the motional degrees of freedom within the optical traps, particularly along the axial direction. Thus, the environment can be modeled as a harmonic oscillator of frequency $\omega$, with free Hamiltonian $H_E^{(0)} = \omega(a_1^\dag a_1 + a_2^\dag a_2 + 1)$, as in Eq.~\eqref{H_mot}. The most relevant system-environment coupling Hamiltonian, $H^{(1)} = H^{(1)}_S \otimes H^{(1)}_E$, as identified above, corresponds to the second-order term arising from axial motion, given in Eq.~\eqref{H_0_delta_R}. In this case, $H^{(1)}_S = \sigma_+^{(1)} \sigma_-^{(2)} + \sigma_-^{(1)} \sigma_+^{(2)}$, and $H^{(1)}_E = -3(x_1 - x_2)^2 / R^2$. The energy scale of the system is set by the dipole-dipole interaction strength $J$, which for polar molecules is typically on the order of $1$~kHz, whereas the energy scale of the axial motion is determined by the trapping frequency $\omega$, which is usually an order of magnitude larger than $J$. Therefore, the adiabatic approximation is justified.\\
To verify this result, Fig.~\hyperref[fig:motion_omega]{\ref{fig:motion_omega}(b)} shows the infidelity $1-F$ of a robust pulse [see Sec.\ref{sect:results}] compared to that of the native gate (i.e., without external controls), plotted as a function of the trapping frequency $\omega$. The simulation includes both the molecular system and the environment, accounting for up to 7 motional states, and uses the experimental parameters reported in Ref.~\cite{Picard_2024_arXiv}. Both the robust pulse and the native gate exhibit improved performance as the trapping frequency $\omega$ increases. This can be intuitively understood by thinking that increasing $\omega$ increases the probability of keeping the molecule in the motional ground state of its harmonic trap, thus reducing the position fluctuations. The robust pulse leads to an improvement in infidelity by one to two orders of magnitude. This improvement is more pronounced at higher trapping frequencies -- for instance, around two orders of magnitude at $\omega \approx 14J$ -- where the adiabatic approximation ($J \ll \omega$) is better satisfied. Conversely, the improvement diminishes at lower trapping frequencies, reducing to about one order of magnitude at $\omega \approx 5J$.

\begin{figure*}
    \centering
    \includegraphics[width=0.98\linewidth]{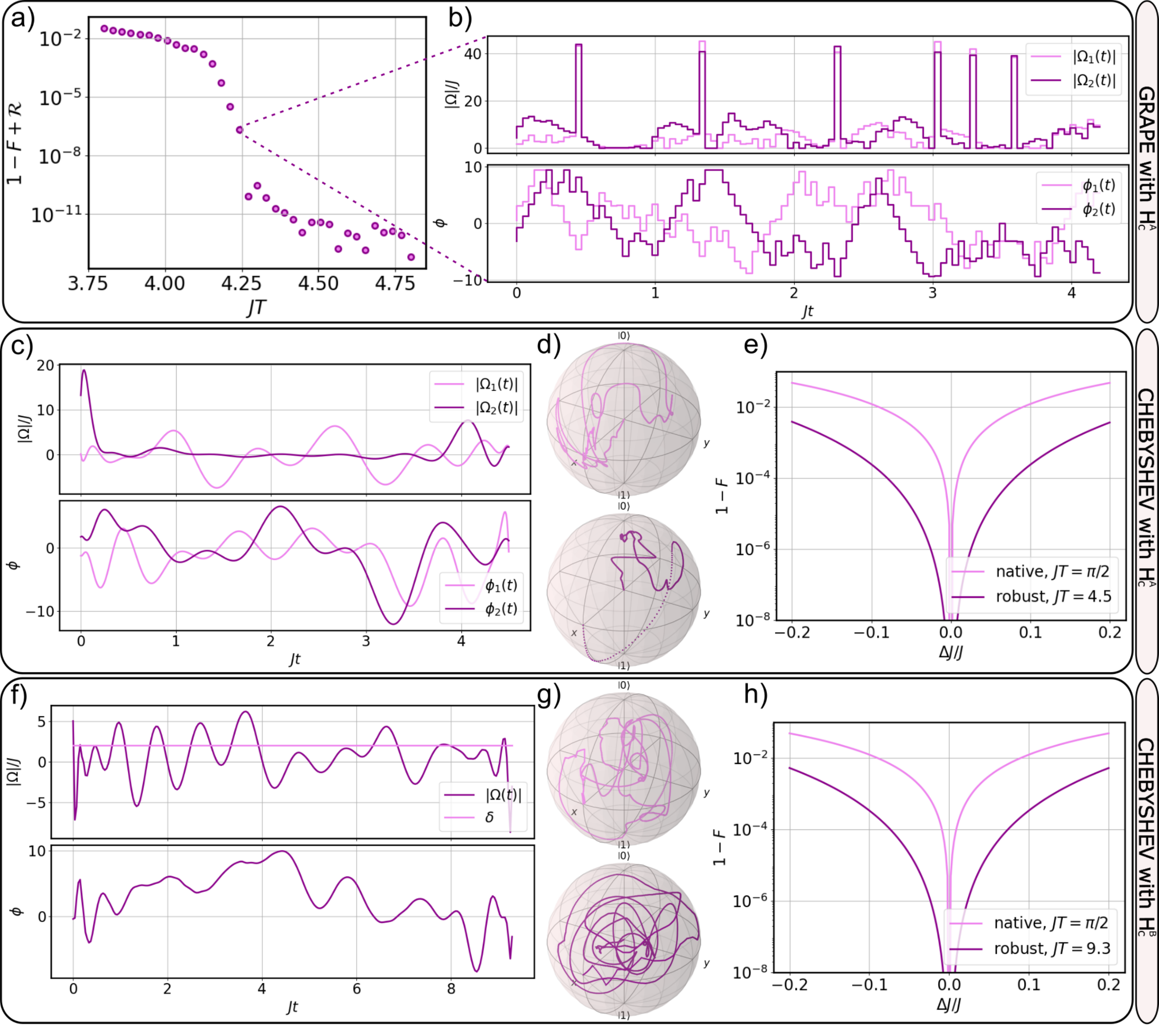}
    \caption{Robust iSWAP gate with respect to noise in the interaction strength $J$ with local controls optimized by GRAPE algorithm (upper rectangle) and Chebyshev polynomials (lower rectangles). \textbf{a)} Cost $\mathcal{C}=1 - F + \mathcal{R}$ found by GRAPE as a function of the dimensionless gate duration $JT$ with the control Hamiltonian $H^A_c$ in Eq.~\eqref{H_c1} (full single-site addressability). Each point is obtained starting with random initial conditions, a discretization of $N=90$ steps and imposing a boundary over the laser amplitude $|\Omega_j|^i<|\Omega_{\text{max}}|=50J$. Changing these parameters we do not observe any significant decrease of the critical gate duration $JT^*\simeq4.2$. \textbf{b)} Amplitude $|\Omega_{1,2}(t)|$ and phase $\phi_{1,2}(t)$ of the shortest GRAPE pulse realizing a robust iSWAP with the control Hamiltonian $H^A_c$ in Eq.~\eqref{H_c1}. The duration of the pulse is $JT^* = 4.2$. Due to numerical instabilities the curves are not smooth. \textbf{c)} Amplitude $|\Omega_{1,2}(t)|$ and phase $\phi_{1,2}(t)$ of the Chebyshev pulse with the control Hamiltonian $H_c^A$ in Eq.~\eqref{H_c1} (full single-site addressability). Negative values of the amplitude $|\Omega_{1,2}(t)|$ can be turned in positive by simply adding a $\pi$ to the corresponding phase. The duration of the pulse is $JT = 4.5$. \textbf{d)} Evolutions in the Bloch sphere of the reduced density matrix of each qubit (qubit 1 up, qubit 2 below) along the duration of the Chebyshev pulse in Fig. \hyperref[fig:big_pic]{\ref{fig:big_pic}(c)}. The system starts in the separable state $\ket{++}$ on the surface of the sphere, the final state is a maximally entangled state that is represented by the point at the center of the sphere. \textbf{e)} Infidelity $1-F$ of the time-optimal non-robust native gate (light line) and the Chebyshev robust pulse in Fig. \hyperref[fig:big_pic]{\ref{fig:big_pic}(c)} (dark line) as a function of quasi-static relative variations $\Delta J /J$ of the coupling strength. For variations of 10\% the infidelity is still very small $1-F=2\times10^{-4}$, more than two order of magnitude smaller than in the native protocol. \textbf{f)} Amplitude $|\Omega(t)|$, phase $\phi(t)$ and local detuning $\delta$ of the Chebyshev local pulse with the control Hamiltonian $H_c^B$ in Eq.~\eqref{H_c2} (global microwave and local detuning). The duration of the pulse is $JT = 9.3$. \textbf{g)} Evolutions in the Bloch sphere of the reduced density matrix of each qubit (qubit 1 up, qubit 2 below) along the duration of Chebyshev pulse in Fig. \hyperref[fig:big_pic]{\ref{fig:big_pic}(f)}. \textbf{h)} Infidelity $1-F$ of the time-optimal non-robust native gate (light line) and the Chebyshev robust pulse in Fig. \hyperref[fig:big_pic]{\ref{fig:big_pic}(f)} (dark line) as a function of quasi-static relative variations $\Delta J /J$ of the coupling strength.}
    \label{fig:big_pic}
\end{figure*}

\section{\label{section_3}Numerical methods}

In this section, we preset two numerical approaches to shape the control parameters in order to obtain robust gates. In Sec.~\ref{sect:grape} we describe the GRAPE algorithm, a gradient ascent method that allows one to achieve robust and time optimal gates. The GRAPE's optimal pulses are not guaranteed to converge to curves that are smooth functions of time. Thus in Sec.~\ref{sect:cheby}, we present a method to optimize a linear combination of Chebyshev polynomials, that guarantees smooth pulse behavior, which is usually more desirable for experimental implementations.

\subsection{\label{sect:grape}GRAPE algorithm}

The GRAPE (Gradient Ascent Pulse Engineering) algorithm is a well established optimal control technique that, given an initial state $\ket{\psi(0)}$, a pulse duration $T$ and an Hamiltonian $H(\vec{u}(t))$ depending on some controls $\vec{u}(t)$, allows one to find the optimal controls minimizing an arbitrary cost functional $\mathcal{C}(\ket{\psi(T)})$ \cite{Khaneja_2005}. To reach the minimum of the cost functional, GRAPE uses the gradient descent method. Here, we employ the BFGS (Broyden-Fletcher-Goldfarb-Shanno) method \cite{SciPy_2020}.\\
To find the time-dependent control parameters, a piecewise constant ansatz is adopted, dividing the pulse duration $T$ in $N$ time steps of duration $\Delta t = T/N$ and assuming that within each time-step each control has a constant value $\vec{u}(t) = \vec{u}^i$ with $t\in[i\Delta t, (i+1)\Delta t]$ and $i=0,1,...,N-1$.\\
Assuming full single-site addressability as in the control Hamiltonian $H_c^A(\Omega_{1,2}(t))$ in Eq.~\eqref{H_c1}, the control functions to be optimized are the time-dependent Rabi frequency amplitudes $|\Omega_j(t)|$ and phases $\phi_j(t)$ for each qubit $j=1,2$. Six additional parameters can also be introduced to account for the fact that the target gate $U_0(\bar{T})$ has to be reproduced up to two independent single-qubit operations $R$ (whose error cost is typically much smaller than that of entangling gates), each parameterized by three angles $\theta$, $\varphi$ and $\lambda$ \cite{Nielsen_Chuang_2010}
\begin{align}
    \nonumber
    R(\theta, \varphi, \lambda) & =
    \begin{pmatrix}
        \cos(\theta/2) & -e^{i\lambda} \sin(\theta/2)\\
        e^{i\varphi}\sin(\theta/2) & e^{i(\varphi+\lambda) \cos(\theta/2)}
    \end{pmatrix}\\
    \label{R_operation}
    & = e^{\frac{i}{2}(\varphi+\lambda)}R_z(\varphi)R_y(\theta)R_z(\lambda).
\end{align}
In this way the full set of parameters is $\vec{u}^A=\{|\Omega_j|^i,\phi_{j}^i,\theta_j,\varphi_j,\lambda_j\}^{i=0,1,...,N-1}_{j=1,2}$.\\

One can also optimize the global drive $\Omega(t)$, in the control Hamiltonian $H_c^B(\Omega(t))$ in Eq.~\eqref{H_c2}, applying at the end a global single-qubit rotation $R$ \eqref{R_operation}. Thus, in this second case, the the full set of parameters to optimize is $\vec{u}^B=\{|\Omega|^i,\phi^i,\theta,\varphi,\lambda\}^{i=0,1,...,N-1}$.\\
In order to obtain a pulse that implements the desired gate while ensuring robustness, we choose the following cost functional $\mathcal{C}$\cite{Jandura_2023}
\begin{equation}
    \label{C}
    \mathcal{C}(\ket{\psi(T)}) = 1 - F + \mathcal{R},
\end{equation}
where $\mathcal{R}$ is the robustness, as defined in Eq.~\eqref{R}, while $F$ is the Bell state fidelity \cite{Theis_2016, Levine_2019, Graham_2019, Robicheaux_2021}
\begin{equation}
    \label{F}
    F = \frac{1}{16}\left| \sum_{q\in Q} \braket{\psi_q^{(0)}(T)| R^{\otimes2} U_0(\bar{T})|q} \right|^2.
\end{equation}
The latter measures the fidelity between the zeroth-order state $\ket{\psi_q^{(0)}(T)}$, obtained by applying the pulse to the state $\ket{q}$ of the computational basis $Q$, and the corresponding desired output state $U_0(\bar{T})\ket{q}$, up to the two single-qubit operations $R$ \eqref{R_operation}. Minimizing the objective cost $\mathcal{C}$ in Eq.~\eqref{C}, particularly by driving it to zero, ensures that the optimal pulse reproduces the target iSWAP gate $U_0(\bar{T})$ exactly and that it is insensitive to first-order variations in the interaction strength $J$. Note that it is also possible to include in the cost functional $\mathcal{C}$ in Eq.~\eqref{C} additional robustness terms related to uncertainties in other parameters. For instance, in our case, one could design a pulse that, in addition to being robust to noise in $J$, is also robust to noise in the amplitude $|\Omega|$ or the detuning $\delta$.

\subsection{\label{sect:cheby}Optimization with Chebyshev polynomials}

Although it was empirically shown that GRAPE algorithm can converge to a smooth curve when optimizing for the time-optimal gate duration \cite{Jandura_2022, Jandura_2023}, there are no guarantees that this will happen, e.g. due to numerical instabilities. To solve this issue a possible strategy is to include in the cost functional in Eq.~\eqref{C} a regularizer that penalizes discontinuous solutions \cite{Giudici_2024}, or optimizing for slightly longer pulses and using as initial condition a smooth pulse instead of random numbers.\\
Another way to design the controls, maybe more naive but that in our case performs almost as well as the GRAPE approach, is decomposing the control functions in terms of Chebyshev polynomials \cite{Ma_2023}. The latter form a complete basis set of functions \cite{Rivlin_2020} and guarantee that the final pulse is smooth. Chebyshev polynomials of the first kind $T_n$ form an orthonormal basis, so a generic piecewise smooth and continuous function $f(x)$ defined on the domain $-1\leq x \leq 1$ can be expanded as
\begin{equation}
    \label{Cheby_exp}
    f(x) = \sum_{n=0}^\infty c_n T_n(x).
\end{equation}
Adjusting the domain in the right time interval $[0,T]$ and truncating the expansion in Eq.~\eqref{Cheby_exp} at a finite order $M$, allows one to write the controls as $|\Omega_j(t)|= \sum_{n=0}^{M} a_j^n T_n(t)$ and $\phi_j(t)= \sum_{n=0}^{M} b_j^n T_n(t)$. In this way the full set of parameters to be optimezed becomes $\vec{u}^A=\{a_j^n, b_j^n, \theta_j, \varphi_j, \lambda_j\}_{j=1,2}^{n=0,1,...,M}$ for the control Hamiltonian $H_c^A$ in Eq.~\eqref{H_c1}, and $\vec{u}^B=\{a^n, b^n, \theta, \varphi, \lambda\}^{n=0,1,...,M}$ for the control Hamiltonian $H_c^B$ in Eq.~\eqref{H_c2}. The optimal parameters are the ones that minimize the cost functional $\mathcal{C}$ in Eq.~\eqref{C}. Once a smooth pulse is found, it is also possible to use it as an initial condition for GRAPE to further decrease infidelity. This often allows one to remain close to the smooth solution.

\section{\label{sect:results}Results: The robust and time optimal iSWAP gate}

In this section we present the pulses to realize robust iSWAP gates and a Bell state preparation. In Sec.~\ref{sect:full-single-site-address}, we show the results of the optimization with full single-site addressability controls. In Sec.~\ref{sec:local_detuning}, we show the results of the optimization with local detuning, and finally in Sec.~\ref{sect:state_preparation}, we discuss a robust state preparation with fully global controls. In all these cases we report a decrease in infidelity $1-F$ by two orders of magnitude in the presence of 10\% variations $\Delta J$, with respect to the non-robust protocols.

\subsection{\label{sect:full-single-site-address}Full single-site addressability \texorpdfstring{$H_c^A$}{}}

We first consider the case of local controls $H_c^A$ in Eq.~\eqref{H_c1}, with full single-site addressability. Using GRAPE, we search for robust pulses starting from a gate duration of $T=\bar{T}=\pi/2$, where we observe that the infidelity and the robustness cannot vanish simultaneously . As $T$ is gradually increased, the minimum gate cost function $\mathcal{C}$ drops, finally reaching the numerical zero (set at $10^{-10}$) around $JT^*\sim4.2$, as shown in Fig. \hyperref[fig:big_pic]{\ref{fig:big_pic}(a)}. For gate duration $T>T^*$ the cost remains zero. A similar behavior has recently been observed in \cite{Jandura_2022} for CZ gates with Rydberg atoms. Starting with random initial conditions, pulses in the interval $T>T^*$ are often  found to be not smooth, whereas shorter pulses with $T<T^*$ tend to be smoother. The critical time $T^*$ typically represents not only the shortest pulse that achieves zero cost function $\mathcal{C}$, but also the longest pulse exhibiting smooth behavior as a function of time. However, in our case, due to numerical instabilities the pulse obtained from GRAPE at the critical gate duration $T^*$ exhibits discontinuities and non-smooth behavior, as shown in Fig. \hyperref[fig:big_pic]{\ref{fig:big_pic}(b)}, making experimental reproduction very challenging. For this reason, we employ Chebyshev polynomials, to provide a smoother solution at the cost of a slightly longer pulse. Using this second approach and with just $M=20$ polynomials  we find a smooth pulse of duration $T = 4.5/J$, which is only 7\% longer than the GRAPE one, but much smoother, and thus probably easier to be reproduced in an experiment, as shown in Fig. \hyperref[fig:big_pic]{\ref{fig:big_pic}(c)}. Slightly shorter pulses can be designed increasing the order $M$ of Chebyshev polynomials expansion, at the cost of a more irregular shape, whereas we can attain even smoother pulses reducing this same number, this time at the cost of longer gate duration.\\

As shown in Fig. \hyperref[fig:big_pic]{\ref{fig:big_pic}(e)} we analyze the performance of the optimal robust pulse looking at the infidelity $1-F$ as a function of the quasi-static variations of the coupling strength $\Delta J$ and comparing it to the results of the native gate, where by native we refer to the protocol in which the two qubits interact for a time $\bar{T}=\pi/(2J)$ without any additional control field. In the presence of 10\% variations $\Delta J / J$, the robust pulse achieves a fidelity of $F > 0.9997$, representing an improvement of two orders of magnitude over the time-optimal gate, where no control is applied. The fidelity remains above $F > 0.996$ even under 20\% variations.

\begin{figure*}[ht]
    \centering
    \includegraphics[width=0.98\linewidth]{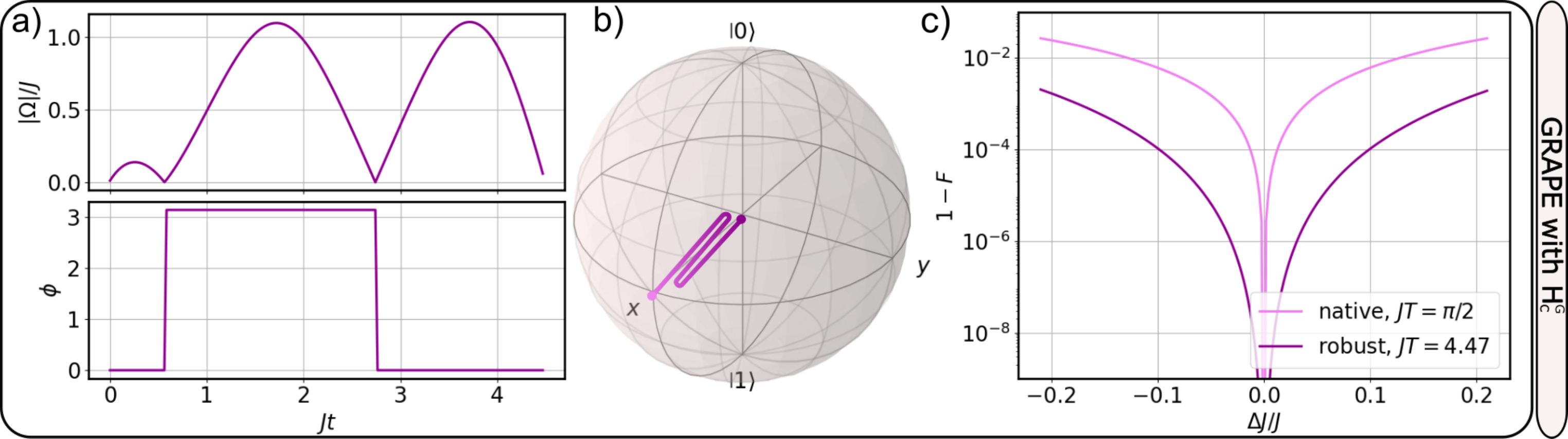}
    \caption{Protocol to prepare a maximally entangled state robustly with respect to noise in the interaction strength $J$, using a global control Hamiltonian $H_c^G$ in Eq.~\eqref{H_c}. \textbf{a)} Amplitude $|\Omega(t)|$ (in units of $J)$ and phase $\phi(t)$ of the global robust pulse for the state preparation. The duration of the pulse is $JT=4.47$. \textbf{b)} The evolution of the reduced density matrix of one qubit along the duration of the pulse, plotted on the Bloch sphere. The system starts in the state $\ket{++}$, which is separable and then the reduced density matrix of each one of the two qubits is represented by a point on the surface of the sphere. Since the initial state is symmetric and the pulse global, the evolution of the states of the two qubits is identical. The final state is a maximally entangled state with purity 0.5, which is represented by the point at the center of the sphere. The motion happens all on the $x$ axis of the Bloch sphere, but to visualize it better the trajectory has been stretched. \textbf{c)} Performance of the pulse with respect to a time-optimal non-robust state preparation, in which the two qubits are let interact for a time $T=\pi/(2J)$, without any additional control field. For static deviations $\Delta J/J$ of 10\%  of the coupling strength $J$ the infidelity is still very small $1-F=10^{-4}$, 60 times smaller than in the native protocol. }
    \label{fig:state_prep}
\end{figure*}

\subsection{\label{sec:local_detuning}Local detuning \texorpdfstring{$H_c^B$}{}}

The results of the pulse in Fig. \hyperref[fig:big_pic]{\ref{fig:big_pic}(c)}, discussed in Sec.~\ref{sect:full-single-site-address}, are promising; however, they require full single-site addressability. This may not be possible in certain physical setups, e.g. polar molecules driven by microwave fields. For this reason, we also consider the controls $H_c^B$ in Eq.~\eqref{H_c2}, driven by a global control field only and with a constant local detuning fixed at $\delta=2J$. In general, larger values of the local detuning lead to shorter time-optimal robust pulses, but starting from around $\delta = 2J$, this speed improvement becomes increasingly marginal and comes at the cost of greater experimental complexity.\\

Using the GRAPE algorithm, we see that a robust time-optimal pulse can be obtained also in this case, for a duration duration $JT^*\sim8$, this is significantly longer than what found with full single-site addressability, in Sec.~\ref{sect:full-single-site-address}. As before, the GRAPE pulse at the critical gate duration $T^*$ exhibits discontinuities and non-smooth behavior. Therefore, we choose to optimize a Chebyshev polynomial decomposition up to order $M=30$. In this way a robust pulse of duration $T=9.3/J$ is found, which is 14\% longer than the time-optimal but with a much smoother shape as shown in Fig. \hyperref[fig:big_pic]{\ref{fig:big_pic}(f)}. It is possible to obtain slightly shorter pulses by increasing the order $M$. Another possibility is increasing the local detuning $\delta$. As shown in Fig. \hyperref[fig:big_pic]{\ref{fig:big_pic}(h)} we analyze the performance of the optimal robust pulse with regard to the infidelity $1-F$ as a function of the variations of the coupling strength $\Delta J$ and comparing it with the results of the native gate with no control fields. We observe a reduction in infidelity by two orders of magnitude in the presence of 10\% variations $\Delta J / J$, achieving a fidelity of $F > 0.9996$, while maintaining a fidelity of $F > 0.994$ under 20\% variations, thus similar to the case with full control shown above in Sec.~\ref{sect:full-single-site-address}. This constitutes one of the main result of this work, paving the way to the implementation of high fidelity entangling gate, as the iSWAP gate with polar molecules, even in the presence of significant noise in the coupling strength.\\

In a system of polar molecules, the main source of noise affecting the interaction strength is the motion along the axial direction, as discussed above in Sec.~\ref{sec:axial_mot}. In Eq.~\eqref{Delta_J_mot}, the magnitude of this noise, $\Delta J$, was estimated using the numerical parameters from the experiment in Ref.~\cite{Picard_2024_arXiv}, yielding $\Delta J / J \approx 7\%$. In this scenario, the adiabatic approximation is satisfied, since the trapping frequency is larger than the interaction strength, in particular $\omega \approx 7J$. Given this assumption and the results from Sec.~\ref{sec:noise-coupling_env}, we know that the robustness of the previously designed pulses -- originally tailored for fluctuations in Hamiltonian parameters -- is also guaranteed in the presence of coupling to motion. In Fig.~\hyperref[fig:motion_omega]{\ref{fig:motion_omega}(b)} the infidelity $1-F$ of the robust pulse is plotted as function of the trapping frequency $\omega$, and it is compared with the native gate with no control fields. The simulation includes the coupling to the harmonic trapping potential, and uses the experimental parameters reported in Ref.~\cite{Picard_2024_arXiv}.The robust pulse yields an improvement in infidelity of approximately one order of magnitude. In particular, at $\omega \approx 7J$, the fidelity exceeds $F > 0.9995$, which aligns well with the expected behavior under quasi-static fluctuations of the interaction strength of around 10\%, as discussed above. The fidelity and the improvement provided by the robust pulse become even more pronounced at higher trapping frequencies, where the adiabatic approximation is more accurately satisfied. This result demonstrates that high-fidelity two-qubit gates with molecules are achievable within the reach of current experimental capabilities.

\subsection{\label{sect:state_preparation}Robust Bell state preparation with global control \texorpdfstring{$H_c^G$}{}}

In the Sec.~\ref{sect:results} above, we discussed the construction of a robust entangling iSWAP gate by applying local controls. However, if one is instead interested in robustly preparing a single entangled state, such as a Bell state, from a given initial state, without concern for the robustness of the evolution of all other states in the Hilbert space, this can be achieved using a global controls only. In Sec.~\ref{sec:robust?_no_sorry}, the need for a local control was motivated based on criterion \ref{criterion_iii} of Sec.~\ref{sect:criteria}. The reason the criterion \ref{criterion_iii} does not apply in the state preparation problem is that, when considering the evolution of a single initial state, there is no sum over all states $\ket{q}$ in a basis $Q$ in Eq.~\eqref{comp_2}, thus it does not become a trace.\\

We verified this by designing a global pulse to prepare a maximally entangled state from the initial state $\ket{++} = (\ket{00} + \ket{01} + \ket{10} + \ket{11})/2$ using the GRAPE algorithm discussed in Sec.~\ref{sect:grape}. The results are shown in Fig. \ref{fig:state_prep}. As for the gate's pulse (see Fig. \hyperref[fig:big_pic]{\ref{fig:big_pic}(a)}) we determine the critical time $T^*=4.47/J$, thus the minimum gate duration that allows GRAPE to find a pulse with zero cost $\mathcal{C}$. Contrary to what seen for the gate in Secs. \ref{sect:full-single-site-address} and \ref{sec:local_detuning} above, the pulse with gate duration $T^*$ has a simple and smooth behavior, as shown in Fig. \hyperref[fig:state_prep]{\ref{fig:state_prep}(a)}. This is due to the fact that a single state preparation is a much simpler protocol that requires less constraints on the optimal controls parameters. In Fig. \hyperref[fig:state_prep]{\ref{fig:state_prep}(c)}, the infidelity $1-F$ is plotted as a function of the relative variations of the coupling strength $\Delta J/J$ for both the robust pulse and the time-optimal native state preparation. The infidelity is decreased by almost two orders of magnitude in the presence of 10\% variations $\Delta J / J$, with a fidelity of $F > 0.9998$, while maintaining a fidelity of $F > 0.998$ under 20\% variations.

\section{Conclusions and outlook}

In this work, we have presented a theoretical framework leveraging quantum optimal control methods to address a critical challenge in quantum computing: the sensitivity of two-qubit entangling gates to variations in experimental parameters. We derived a set of criteria to determine whether robust gate sequences can be designed for a given system and control Hamiltonians. We also demonstrated that these criteria apply not only to fluctuations of system or control parameters, but also to the more general case of coupling to a noisy environment, e.g. thermal motion in optical traps, provided the adiabatic approximation holds.\\
Focusing on systems governed by an exchange interaction Hamiltonian and realizing an iSWAP gate, we found that achieving robustness to variations in interaction strength is not possible with global control schemes. However, by breaking the exchange symmetry -- either with full single-site addressability controls or just with a local detuning -- we were able to develop robust protocols, yielding smooth pulses with high fidelity $F>0.9996$, even with a 10\% fluctuation of the coupling strength. This result is further confirmed by simulations that include the coupling to the harmonic motion of the molecule within the trap. These findings have significant implications for physical platforms such as ultracold polar molecules, where the exchange Hamiltonian is implemented by the dipole-dipole interaction and the variability of the interaction strength, mainly due to motions in the traps, remains a significant source of error. Our results show that within reach of current experiments it is possible to achieve high-fidelity two-qubit gates with molecules. For the considered noise sources, this could lead experiments to potentially surpass the error probability threshold that allows for quantum error correction to be possible. \\

A possible future development of our work is the application of similar protocols to neutral atoms. Neutral atoms are already one of the most promising platforms for quantum computing: quantum information can be encoded in the internal states of atoms, while a possible mechanism to realize fast two-qubit entangling gates is through strong resonant electric dipole interaction between electronically highly excited Rydberg states, which enables the coherent exchange of energy between nearby atoms prepared in different Rydberg states on micro- or even nanosecond timescales. However, as in polar molecules, the main challenge is the sensitivity to variations of the coupling strength. The intense interaction induces significant motion in the two atoms, thus making it important to get robust pulses with a full quantum description of the atomic motion in the harmonic potential of the optical traps, as discussed in this work.

\appendix
\section{\label{sect:ext_robustness}Extended robustness}

In Sec.~\ref{sect:robustness}, we define a gate as robust when the first-order correction to the state due to the noisy perturbation vanishes. In a more precise sense one can say that a quantum gate is robust either by sending to zero the norm of the first-order state $\ket{\psi_q^{(1)}(T)}$, thus the robustness $\mathcal{R}$ in Eq.~\eqref{R}, or by sending to zero the norm of the following state
\begin{equation}
    \label{phi_1}
    \ket{\varphi_q^{(1)}(T)} = \ket{\psi_q^{(1)}(T)} - i\alpha\ket{\psi_q^{(0)}(T)}.
\end{equation}
In this second and more general case, the state $\ket{\psi_q^{(1)}(T)}$ is allowed to have non-zero norm, but only if it is parallel to the intended state $\ket{\psi_q^{(0)}(T)}$, up to a phase. The parameter $\alpha\in\mathbb{R}$ takes into account the degree of orthogonality between $\ket{\psi_q^{(0)}(T)}$ and $\ket{\psi_q^{(1)}(T)}$, i.e. $\alpha=|\braket{\psi_q^{(0)}|\psi_q^{(1)}}|$. The extended robustness $\mathcal{R}_e$ can be defined as the sum of the norms of the latter states
\begin{equation}
    \label{R_e}
    \mathcal{R}_e = \sum_{q\in Q} \braket{\varphi_q^{(1)}(T)|\varphi_q^{(1)}(T)}.
\end{equation}
Inserting the expression in Eq.~\eqref{phi_1} for $\ket{\varphi_q^{(1)}(T)}$ in the definition of the extended robustness $\mathcal{R}_e$ in Eq.~\eqref{R_e}, one obtains
\begin{align}
    \mathcal{R}_e = & \sum_{q\in Q} \braket{\psi_q^{(1)}(T)|\psi_q^{(1)}(T)} + \alpha^2 \sum_{q\in Q} \braket{\psi_q^{(0)}(T)|\psi_q^{(0)}(T)}-\\
    &-2 \alpha \Im\sum_{q\in Q} \braket{\psi_q^{(0)}(T)|\psi_q^{(1)}(T)}\\
    \label{comp_1}
    = & \mathcal{R} + K\alpha^2 -2\alpha \Im{\sum_{q\in Q} \braket{\psi_q^{(0)}(T)|\psi_q^{(1)}(T)}},
\end{align}
where $K$ is the sum of norms of the zeroth-order states $\ket{\psi_q^{(0)}(T)}$, which is a positive number (exactly $4$ if the gate is robust). Looking at the last term in Eq.~\eqref{comp_1} and using the Eqs. \eqref{psi_0} and \eqref{psi_1} for $\ket{\psi_q^{(0)}(T)}$ and $\ket{\psi_q^{(1)}(T)}$ respectively, one can prove that
\begin{align}
    & \sum_{q\in Q} \braket{\psi_q^{(0)}(T)|\psi_q^{(1)}(T)} =\\
    & = -i \sum_{q\in Q} \int_0^T dt \bra{q}U^{(0)}(0,T)^\dag U^{(0)}(t,T) H^{(1)} U^{(0)}(0,t)\ket{q}\\
    & =  -i \int_0^T dt \Tr\left\{U^{(0)}(t,0) H^{(1)} U^{(0)}(0,t)\right\}\\
    \label{comp_3}
    & =  -i T \Tr\left\{H^{(1)}\right\},
\end{align}
where the fact that $Q$ is a basis is used to recover a trace, and the cyclic property of traces is applied to permute the evolution operators. Inserting Eq.~\eqref{comp_3} in Eq.~\eqref{comp_1} one finally obtains the following generalized expression for robustness
\begin{equation}
    \mathcal{R}_e = K\alpha^2 + 2\alpha T \Tr\left\{H^{(1)}\right\} + \mathcal{R}.
\end{equation}
If the first-order Hamiltonian is traceless $\Tr\{H^{(1)}\}=0$, then the robustness requirement $\mathcal{R}_e=0$ is equivalent to $\mathcal{R}=0$. Notice that $H^{(1)}$ can always be made traceless by performing an unitary transformation, i.e. by adding to it an identity matrix multiplied by $-\Tr\{H^{(1)}\}/\dim Q$. The term proportional to the identity does not contribute to the dynamics besides an irrelevant overall global phase. Therefore, up to a unitary transformation, one can neglect the extended robustness $\mathcal{R}_e$ \eqref{R_e} and focus on sending the robustness  $\mathcal{R}$ \eqref{R} to zero.

\section{\label{sect:noise}Quasi-static noise model}

\subsection{Quasi-static noise}

In a quasi-static noise model the noise in the coupling strength is assumed to be in the form $J(t)=J+\Delta J$, where $\Delta J$ is an unknown and constant (or slowly varying on the timescale of the gate) deviation from the expected value $J$. The value of the deviation $\Delta J$ can vary from one experiment to the next, but it does not change significantly during a single experiment. The value of the deviation $\Delta J$ is considered to be a random variable that follows a normal distribution with mean $0$ and standard deviation $\sigma_J$, such that $\Delta J \sim \mathcal{N}(0,\sigma_J^2)$.\\

\begin{figure}[ht]
    \centering
    \includegraphics[width=\linewidth]{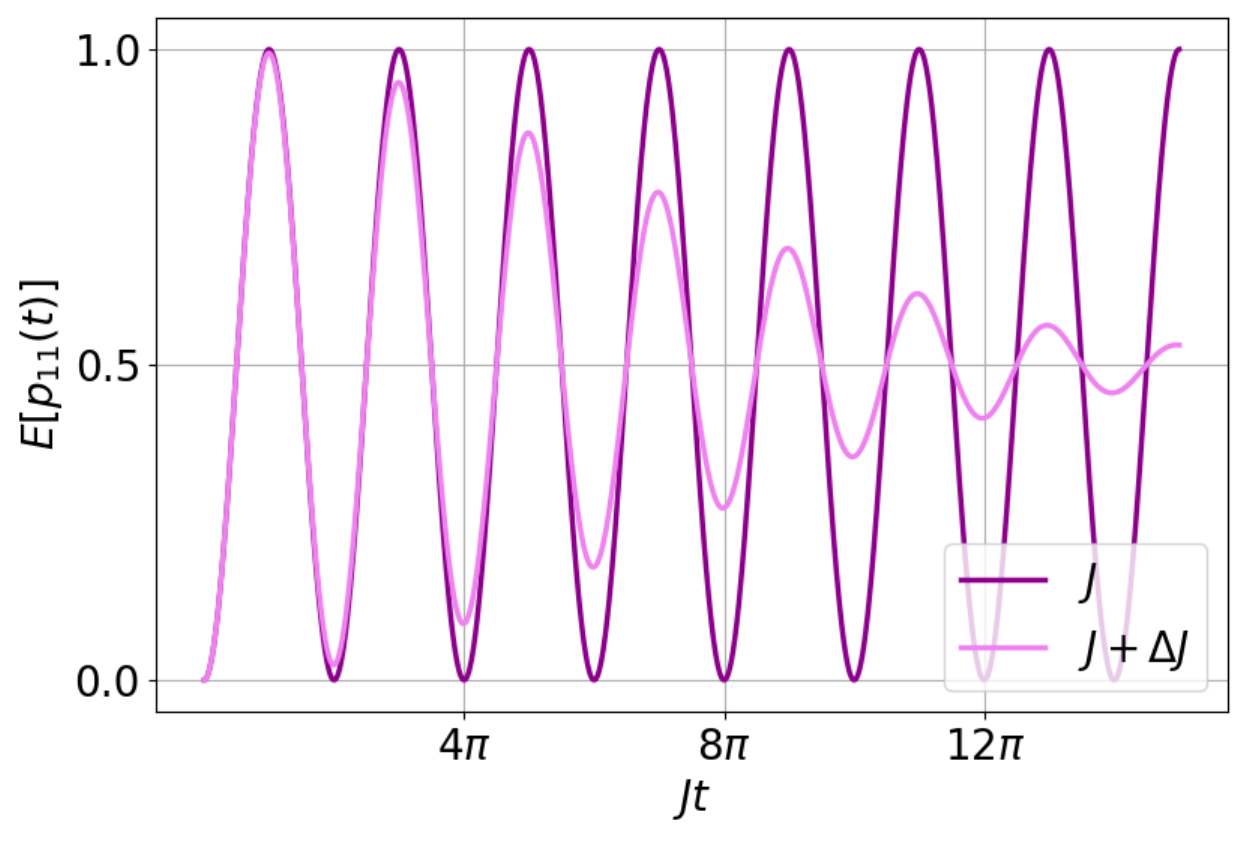}
    \caption{Expectation value of population of the $\ket{11}$ state as a function of the interacting time $t$ for a pair of qubits undergoing dipole-dipole exchange, according to the evolution operator in Eq.~\eqref{U_hop}. Time $t$ is expressed in $J$ units. The dark line represents the case in which there is no noise in the system, with coherent oscillations of the population. The light line represents the case of a quasi-static noise $\Delta J$, randomly sampled from a central Gaussian distribution with standard deviation $\sigma_J=0.05J$, that gives rise to decoherence.}
    \label{plot_noises}
\end{figure}

\subsection{Effects of the noise}

To evaluate the effect of the quasi-static noise model on the qubit system, the preparation of a maximally entangled state is considered. This operation is much simpler than implementing a full entangling gate, but it serves as an instructive case study to observe how the noise model affects the final outcome of quantum state manipulation. Following the procedure of a classical Ramsey interferometry experiment, the process begins from the state $\ket{00}$, followed by the application of a global $\pi/2$ pulse. The two qubits then interact for a time $t$ according to $U_0(t)$ in Eq.~\eqref{U_hop}, and finally, a second global $\pi/2$ pulse is applied. In this way
\begin{align}
    & \ket{00} \xrightarrow{\pi/2} \ket{++}\\
     & \xrightarrow{U_0(t)} \frac{1}{2}\left(\ket{00} + e^{-i\Theta(t)} (\ket{01}+\ket{10})+ \ket{11}\right)\\
    \label{Ramsey_seq}
    & \xrightarrow{\pi/2} \frac{1}{2}\left( \ket{00} (1+e^{-\Theta(t)}) + \ket{11} (1-e^{-i\Theta(t)})\right),
\end{align}
where $\ket{+}=(\ket{0}+\ket{1})/\sqrt{2}$, and
\begin{align}
    \label{Theta}
    \Theta(t) = & \int_0^tdt' J(t')\\
    = &
    \begin{cases}
        Jt & \text{no noise}\\
        (J+\Delta J)t & \text{quasi-static noise} \\
    \end{cases}
\end{align}
The population $p_{11}(t)$ of the state $\ket{11}$ at the end of the Ramsey sequence \eqref{Ramsey_seq} with an interaction time $t$, reads
\begin{equation}
    \label{p_11}
    p_{11}(t) = \frac{1}{2}(1-\cos\Theta(t)).
\end{equation}
In the case of zero noise one observes coherent oscillations of the population between 0 and 1 with frequency $J$, described by $(1 - \cos(Jt))/2$, as shown in Fig. \ref{plot_noises}. In the case of quasi-static noise the expectation value of the expression in Eq.~\eqref{p_11} can be evaluated using the corresponding $\Theta$ in Eq.~\eqref{Theta} and a normal distribution $\mathcal{N}(0,\sigma_J^2)$
\begin{align}
    E[p_{11}(t)] = & \int_{-\infty}^{\infty} d(\Delta J) \frac{e^{-\Delta J^2/2\sigma_J^2}}{\sqrt{2\pi}\sigma_J}\frac{1}{2}(1-\cos((J+\Delta J)t)\\
    = & \frac{1}{2}\left(1-e^{-t^2\sigma_J^2/2}\cos(Jt)\right).
\end{align}
In this case, decoherence is observed with damped oscillations of the population and a homogeneous dephasing time given by $T_2 = \sqrt{2} / \sigma_J$, as shown in Fig. \ref{plot_noises}, due to the effect of the noise.

\section*{Acknowledgments}

We are grateful to R. R. Riso, K. K. Ni, L. R. B. Picard, G. E. Patenotte, and D. Wellnitz for their insightful discussions and thoughtful feedback, which greatly contributed to the development of this work. This research has received funding from the European Union’s Horizon Europe programme HORIZON-CL4-2021-DIGITAL-EMERGING-01-30 via the project 101070144 (EuRyQa) and from the French National Research Agency under the Investments of the Future Program projects ANR-21-ESRE-0032 (aQCess), ANR-22-CE47-0013-02 (CLIMAQS), and ANR-22-CMAS-0001 France 2030 (QuanTEdu-France).

\bibliography{main}

\end{document}